\documentclass[12pt]{article}

\pdfoutput=1

\usepackage{graphicx}
\usepackage{epsfig,amsfonts,amssymb,upgreek}
\usepackage{amsmath}
\usepackage[margin=2.5cm]{geometry}
\usepackage[colorlinks=true,urlcolor=blue,linkcolor=blue,citecolor=red]{hyperref}
\usepackage{cite}
\usepackage{framed}
\usepackage{young}
\usepackage[margin=2.5cm]{geometry}
\usepackage{breqn}
\usepackage{flexisym}
\usepackage{ifpdf}
\usepackage[makeroom]{cancel}
\usepackage[normalem]{ulem}
\usepackage{xcolor}

\def\figmulti{

\def\JPicScale{0.6}
\ifx\JPicScale\undefined\def\JPicScale{1}\fi
\unitlength \JPicScale mm
\begin{picture}(150,90)(0,0)
\linethickness{0.3mm}
\put(10,10){\line(1,0){140}}
\put(150,10){\vector(1,0){0.12}}
\linethickness{0.3mm}
\put(10,10){\line(0,1){80}}
\put(10,90){\vector(0,1){0.12}}
\linethickness{0.3mm}
\put(10,80){\line(1,0){140}}
\linethickness{0.3mm}
\multiput(80,10)(0,1.97){36}{\line(0,1){0.99}}
\linethickness{0.3mm}
\linethickness{0.3mm}
\multiput(80,10)(0.89,1.78){23}{\multiput(0,0)(0.11,0.22){4}{\line(0,1){0.22}}}%
\linethickness{0.3mm}
\multiput(100,50)(0,1.94){16}{\line(0,1){0.97}}%
\linethickness{0.3mm}
\multiput(60,50)(1.08,-1.62){19}{\multiput(0,0)(0.11,-0.16){5}{\line(0,-1){0.16}}}%
\linethickness{0.3mm}
\multiput(60,50)(0,1.94){16}{\line(0,1){0.97}}%
\put(120,15){\makebox(0,0)[cc]{$\uptau=-1$}}

\put(115,75){\makebox(0,0)[cc]{$\uptau=0$}}

\put(160,10){\makebox(0,0)[cc]{$x$}}

\put(10,95){\makebox(0,0)[cc]{$\uptau$}}

\end{picture}

}

\def\figvennthree{

\def\JPicScale{0.7}
\ifx\JPicScale\undefined\def\JPicScale{1}\fi
\unitlength \JPicScale mm
}

\def\figone{

\def\JPicScale{0.4}
\ifx\JPicScale\undefined\def\JPicScale{1}\fi
\unitlength \JPicScale mm
\begin{picture}(220,90)(0,0)
\linethickness{0.3mm}
\put(30,40){\line(1,0){180}}
\put(210,40){\vector(1,0){0.12}}
\linethickness{0.3mm}
\put(30,30){\line(0,1){60}}
\put(30,90){\vector(0,1){0.12}}
\put(220,35){\makebox(0,0)[cc]{$x$}}

\put(20,90){\makebox(0,0)[cc]{$\uptau$}}

\linethickness{0.3mm}
\put(30,80){\line(1,0){180}}

\put(118,34){\makebox(0,0)[cc]{$r$}}

\put(165,88){\makebox(0,0)[cc]{$\uptau=0$}}

\put(165,34){\makebox(0,0)[cc]{$\uptau=-1$}}

\linethickness{0.3mm}
\multiput(100,40)(0,1.95){21}{\line(0,1){0.98}}
\linethickness{0.3mm}
\multiput(130,40)(0,1.95){21}{\line(0,1){0.98}}
\put(105,50){\makebox(0,0)[cc]{1}}

\put(135,50){\makebox(0,0)[cc]{2}}

\end{picture}

}

\def\figtwo{

\def\JPicScale{0.6}
\ifx\JPicScale\undefined\def\JPicScale{1}\fi
\unitlength \JPicScale mm
\begin{picture}(220,90)(0,0)
\linethickness{0.3mm}
\put(30,40){\line(1,0){180}}
\put(210,40){\vector(1,0){0.12}}
\linethickness{0.3mm}
\put(30,30){\line(0,1){60}}
\put(30,90){\vector(0,1){0.12}}
\put(220,35){\makebox(0,0)[cc]{$x$}}

\put(20,90){\makebox(0,0)[cc]{$\uptau$}}

\linethickness{0.3mm}
\put(30,80){\line(1,0){180}}
\put(165,85){\makebox(0,0)[cc]{$\uptau=0$}}

\put(165,35){\makebox(0,0)[cc]{$\uptau=-1$}}

\linethickness{0.3mm}
\multiput(100,40)(0,1.95){21}{\line(0,1){0.98}}
\linethickness{0.3mm}
\multiput(60,20)(0.12,0.12){333}{\line(1,0){0.12}}
\linethickness{0.3mm}
\multiput(100,60)(0.12,-0.12){333}{\line(1,0){0.12}}
\linethickness{0.3mm}
\multiput(60,25)(0.12,0.12){333}{\line(1,0){0.12}}
\linethickness{0.3mm}
\multiput(100,65)(0.12,-0.12){333}{\line(1,0){0.12}}
\linethickness{0.3mm}
\multiput(100,60)(1.96,0){26}{\line(1,0){0.48}}
\linethickness{0.3mm}
\multiput(100,65)(1.96,0){26}{\line(1,0){0.48}}
\put(160,60){\makebox(0,0)[cc]{$\uptau$}}

\put(165,65){\makebox(0,0)[cc]{$\uptau+\delta\uptau$}}

\linethickness{0.3mm}
\multiput(100,45)(1.96,0){26}{\line(1,0){0.48}}

\put(156,45){\makebox(0,0)[cc]{$\sig$}}

\end{picture}

}

\def\figthree{

\def\JPicScale{0.7}
\ifx\JPicScale\undefined\def\JPicScale{1}\fi
\unitlength \JPicScale mm
\begin{picture}(220,90)(0,0)
\linethickness{0.3mm}
\put(30,40){\line(1,0){180}}
\put(210,40){\vector(1,0){0.12}}
\linethickness{0.3mm}
\put(30,30){\line(0,1){60}}
\put(30,90){\vector(0,1){0.12}}
\put(220,35){\makebox(0,0)[cc]{$x$}}

\put(20,90){\makebox(0,0)[cc]{$\uptau$}}

\linethickness{0.3mm}
\put(195,35){\makebox(0,0)[cc]{$\uptau=-1$}}

\linethickness{0.3mm}
\multiput(100,40)(0,1.95){21}{\line(0,1){0.98}}
\linethickness{0.3mm}
\multiput(60,20)(0.12,0.12){333}{\line(1,0){0.12}}
\linethickness{0.3mm}
\multiput(100,60)(0.12,-0.12){250}{\line(1,0){0.12}}
\linethickness{0.3mm}
\multiput(60,25)(0.12,0.12){333}{\line(1,0){0.12}}
\linethickness{0.3mm}
\multiput(100,65)(0.12,-0.12){270}{\line(1,0){0.12}}
\linethickness{0.3mm}
\multiput(100,60)(1.96,0){26}{\line(1,0){0.98}}
\linethickness{0.3mm}
\multiput(100,65)(1.96,0){26}{\line(1,0){0.98}}
\put(157,60){\makebox(0,0)[cc]{$\uptau$}}

\put(162,65){\makebox(0,0)[cc]{$\uptau+\delta\uptau$}}

\put(102,75){\makebox(0,0)[cc]{1}}

\put(143,75){\makebox(0,0)[cc]{2}}

\put(125,48){\makebox(0,0)[cc]{$r$}}

\linethickness{0.3mm}
\multiput(140,40)(0,1.95){21}{\line(0,1){0.98}}
\linethickness{0.3mm}
\multiput(110,10)(0.12,0.12){250}{\line(1,0){0.12}}
\linethickness{0.3mm}
\multiput(140,40)(0.12,-0.12){250}{\line(1,0){0.12}}
\linethickness{0.3mm}
\multiput(100,45)(1.95,0){21}{\line(1,0){0.98}}
\put(140,45){\vector(1,0){0.12}}
\put(100,45){\vector(-1,0){0.12}}
\end{picture}

}

\def\figfour{

\def\JPicScale{0.5}
\ifx\JPicScale\undefined\def\JPicScale{1}\fi
\unitlength \JPicScale mm
\begin{picture}(220,95)(0,0)
\linethickness{0.3mm}
\put(5,40){\line(1,0){205}}
\put(210,40){\vector(1,0){0.12}}
\linethickness{0.3mm}
\put(30,10){\line(0,1){85}}
\put(30,95){\vector(0,1){0.12}}
\put(220,35){\makebox(0,0)[cc]{$x$}}

\put(20,90){\makebox(0,0)[cc]{$\uptau$}}

\put(195,35){\makebox(0,0)[cc]{$\uptau=-1$}}

\linethickness{0.3mm}
\multiput(100,40)(0,2){28}{\line(0,1){1}}
\linethickness{0.3mm}
\multiput(20,10)(0.12,0.12){667}{\line(1,0){0.12}}
\linethickness{0.3mm}
\multiput(100,90)(0.12,-0.12){667}{\line(1,0){0.12}}
\linethickness{0.3mm}
\multiput(20,15)(0.12,0.12){667}{\line(1,0){0.12}}
\linethickness{0.3mm}
\multiput(100,95)(0.12,-0.12){667}{\line(1,0){0.12}}
\linethickness{0.3mm}
\multiput(100,90)(1.96,0){26}{\line(1,0){0.98}}
\linethickness{0.3mm}
\multiput(100,95)(1.96,0){26}{\line(1,0){0.98}}
\put(155,90){\makebox(0,0)[cc]{$\uptau$}}

\put(162,95){\makebox(0,0)[cc]{$\uptau+\delta\uptau$}}

\linethickness{0.3mm}
\multiput(140,40)(0,2){28}{\line(0,1){1}}
\linethickness{0.3mm}
\multiput(110,10)(0.12,0.12){250}{\line(1,0){0.12}}
\linethickness{0.3mm}
\multiput(140,40)(0.12,-0.12){250}{\line(1,0){0.12}}
\linethickness{0.3mm}
\multiput(100,45)(1.95,0){21}{\line(1,0){0.98}}
\put(140,45){\vector(1,0){0.12}}
\put(100,45){\vector(-1,0){0.12}}
\put(119,50){\makebox(0,0)[cc]{$r$}}

\put(103,70){\makebox(0,0)[cc]{1}}

\put(143,70){\makebox(0,0)[cc]{2}}

\end{picture}

}

\def\figfive{

\def\JPicScale{0.5}
\ifx\JPicScale\undefined\def\JPicScale{1}\fi
\unitlength \JPicScale mm
\begin{picture}(220,95)(0,0)
\linethickness{0.3mm}
\put(10,25){\line(1,0){205}}
\put(215,25){\vector(1,0){0.12}}
\linethickness{0.3mm}
\put(30,10){\line(0,1){85}}
\put(30,95){\vector(0,1){0.12}}
\put(220,25){\makebox(0,0)[cc]{$x$}}

\put(20,90){\makebox(0,0)[cc]{$\uptau$}}

\put(175,30){\makebox(0,0)[cc]{$\uptau=-1$}}

\linethickness{0.3mm}
\multiput(100,25)(0,1.97){36}{\line(0,1){0.99}}
\linethickness{0.3mm}
\multiput(60,0)(0.12,0.12){333}{\line(1,0){0.12}}
\linethickness{0.3mm}
\multiput(100,40)(0.12,-0.12){333}{\line(1,0){0.12}}
\linethickness{0.3mm}
\multiput(60,5)(0.12,0.12){333}{\line(1,0){0.12}}
\linethickness{0.3mm}
\multiput(100,45)(0.12,-0.12){333}{\line(1,0){0.12}}
\linethickness{0.3mm}
\multiput(100,40)(1.96,0){26}{\line(1,0){0.98}}
\linethickness{0.3mm}
\multiput(100,45)(1.96,0){26}{\line(1,0){0.98}}
\put(155,38){\makebox(0,0)[cc]{${\uptau}_1$}}

\put(165,47){\makebox(0,0)[cc]{${\uptau}_1+\delta{\uptau}_1$}}

\linethickness{0.3mm}
\multiput(140,25)(0,1.97){36}{\line(0,1){0.99}}
\linethickness{0.3mm}
\multiput(55,5)(0.12,0.12){708}{\line(0,1){0.12}}
\linethickness{0.3mm}
\multiput(140,90)(0.12,-0.12){667}{\line(1,0){0.12}}
\linethickness{0.3mm}
\multiput(100,55)(1.95,0){21}{\line(1,0){0.98}}
\put(140,55){\vector(1,0){0.12}}
\put(100,55){\vector(-1,0){0.12}}
\put(115,50){\makebox(0,0)[cc]{$r$}}

\put(103,70){\makebox(0,0)[cc]{1}}

\put(143,70){\makebox(0,0)[cc]{2}}

\end{picture}

}

\def\figsix{

\def\JPicScale{0.7}
\ifx\JPicScale\undefined\def\JPicScale{1}\fi
\unitlength \JPicScale mm
\begin{picture}(220,90)(0,0)
\linethickness{0.3mm}
\put(30,20){\line(1,0){180}}
\put(210,20){\vector(1,0){0.12}}
\linethickness{0.3mm}
\put(30,10){\line(0,1){80}}
\put(30,90){\vector(0,1){0.12}}
\put(217,22){\makebox(0,0)[cc]{$x$}}

\put(20,90){\makebox(0,0)[cc]{$\uptau$}}

\linethickness{0.3mm}
\put(195,15){\makebox(0,0)[cc]{$\uptau=-1$}}

\linethickness{0.3mm}
\multiput(100,20)(0,1.95){31}{\line(0,1){0.98}}
\linethickness{0.3mm}
\multiput(60,20)(0.12,0.12){333}{\line(1,0){0.12}}
\linethickness{0.3mm}
\multiput(100,60)(0.12,-0.12){250}{\line(1,0){0.12}}
\linethickness{0.3mm}
\multiput(60,25)(0.12,0.12){333}{\line(1,0){0.12}}
\linethickness{0.3mm}
\multiput(100,65)(0.12,-0.12){270}{\line(1,0){0.12}}
\linethickness{0.3mm}
\multiput(100,60)(1.96,0){26}{\line(1,0){0.98}}
\linethickness{0.3mm}
\multiput(100,65)(1.96,0){26}{\line(1,0){0.98}}
\put(157,60){\makebox(0,0)[cc]{${\uptau}_1$}}

\put(162,65){\makebox(0,0)[cc]{${\uptau}_1+\delta{\uptau}_1$}}

\put(102,75){\makebox(0,0)[cc]{1}}

\put(143,75){\makebox(0,0)[cc]{2}}

\put(125,48){\makebox(0,0)[cc]{$r$}}

\linethickness{0.3mm}
\multiput(140,20)(0,1.95){31}{\line(0,1){0.98}}
\linethickness{0.3mm}
\multiput(110,10)(0.12,0.12){250}{\line(1,0){0.12}}
\linethickness{0.3mm}
\multiput(140,40)(0.12,-0.12){250}{\line(1,0){0.12}}
\linethickness{0.3mm}
\multiput(100,45)(1.95,0){21}{\line(1,0){0.98}}
\put(140,45){\vector(1,0){0.12}}
\put(100,45){\vector(-1,0){0.12}}
\end{picture}

}

\def\figseven{

\def\JPicScale{0.5}
\ifx\JPicScale\undefined\def\JPicScale{1}\fi
\unitlength \JPicScale mm
\begin{picture}(220,95)(0,0)
\linethickness{0.3mm}
\put(10,25){\line(1,0){205}}
\put(215,25){\vector(1,0){0.12}}
\linethickness{0.3mm}
\put(30,10){\line(0,1){105}}
\put(30,115){\vector(0,1){0.12}}
\put(220,25){\makebox(0,0)[cc]{$x$}}

\put(20,90){\makebox(0,0)[cc]{$\uptau$}}

\put(195,30){\makebox(0,0)[cc]{$\uptau=-1$}}

\linethickness{0.3mm}
\multiput(100,25)(0,1.97){36}{\line(0,1){0.99}}
\linethickness{0.3mm}
\multiput(20,10)(0.12,0.12){667}{\line(1,0){0.12}}
\linethickness{0.3mm}
\multiput(100,90)(0.12,-0.12){667}{\line(1,0){0.12}}
\linethickness{0.3mm}
\multiput(20,15)(0.12,0.12){667}{\line(1,0){0.12}}
\linethickness{0.3mm}
\multiput(100,95)(0.12,-0.12){667}{\line(1,0){0.12}}
\linethickness{0.3mm}
\multiput(100,90)(1.96,0){26}{\line(1,0){0.98}}
\linethickness{0.3mm}
\multiput(100,95)(1.96,0){26}{\line(1,0){0.98}}
\put(155,88){\makebox(0,0)[cc]{${\uptau}_1$}}

\put(167,97){\makebox(0,0)[cc]{${\uptau}_1+\delta{\uptau}_1$}}

\linethickness{0.3mm}
\multiput(140,25)(0,1.97){36}{\line(0,1){0.99}}
\linethickness{0.3mm}
\multiput(110,10)(0.12,0.12){250}{\line(1,0){0.12}}
\linethickness{0.3mm}
\multiput(140,40)(0.12,-0.12){250}{\line(1,0){0.12}}
\linethickness{0.3mm}
\multiput(100,45)(1.95,0){21}{\line(1,0){0.98}}
\put(140,45){\vector(1,0){0.12}}
\put(100,45){\vector(-1,0){0.12}}
\put(115,50){\makebox(0,0)[cc]{$r$}}

\put(103,70){\makebox(0,0)[cc]{1}}

\put(143,70){\makebox(0,0)[cc]{2}}

\end{picture}

}

\usepackage{cite}

\def\ZZZ{{\hbox{ Z\kern-1.6mm Z}}}
\def\RRR{{\hbox{ R\kern-2.4mm R}}}
\def\CCC{{\hbox{ C\kern-2.0mm C}}}
\def\zzz{{\hbox{z\kern-1mm z}}}

\newcommand{\qeq}{{\hbox{=\kern-2.3mm ? \kern.5mm }}}
\renewcommand{\qeq}{=}

\newcommand{\OO}{{\cal O}}

\newcommand{\wt}{\widetilde}

\newcommand{\be}{\begin{equation}}
\newcommand{\ee}{\end{equation}}
\newcommand{\ben}{\begin{eqnarray}\displaystyle}
\newcommand{\een}{\end{eqnarray}}

\newcommand{\refb}[1]{(\ref{#1})}
\newcommand{\p}{\partial}
\newcommand{\sectiono}[1]{\section{#1}\setcounter{equation}{0}}

\newcommand{\gsim}{\stackrel{>}{\sim}}
\newcommand{\lsim}{\stackrel{<}{\sim}}

\def\one{{\hbox{ 1\kern-.8mm l}}}
\def\zero{{\hbox{ 0\kern-1.5mm 0}}}

\newcommand{\bea}[1]{\begin{eqnarray}\label{#1} }
\newcommand{\eea}{\end{eqnarray}}

\def\changes#1{{\color{red} #1}}
\def\changed#1{{\color{blue} #1}}
\def\changec#1{{\color{green} #1}}
\def\changem#1{{\color{red} #1}}
\def\changen#1{{\color{blue} #1}}
\def\changep#1{{\color{red} #1}}
\def\changek#1{{\color{red} #1}}

\def\changes#1{{\color{black} #1}}
\def\changed#1{{\color{black} #1}}
\def\changec#1{{\color{black} #1}}
\def\changem#1{{\color{black} #1}}
\def\changen#1{{\color{black} #1}}
\def\changep#1{{\color{black} #1}}
\def\changek#1{{\color{black} #1}}

\definecolor{orange}{rgb}{1,0.5,0}

\def\sig{\changem{\sigma}}

\def\tcv#1{{\color{magenta} #1}}
\def\tcr#1{{\color{red} #1}}
\def\tcb#1{{\color{blue} #1}}
\def\tcg#1{{\color{green} #1}}
\def\tcy#1{{\color{orange} #1}}
\def\tccy#1{{\color{cyan} #1}}

\numberwithin{equation}{section}

\begin{document}

\begin{flushright}
\end{flushright}

\vskip 12pt

\baselineskip 24pt

\begin{center}
{\Large \bf  Surviving in a Metastable de Sitter Space-Time}

\end{center}

\vskip .6cm
\medskip

\vspace*{4.0ex}

\baselineskip=18pt

\centerline{ \rm 
Sitender Pratap Kashyap$^a$, Swapnamay Mondal$^a$, 
Ashoke Sen$^{a,b}$,
Mritunjay Verma$^{a,c}$}

\vspace*{4.0ex}

\centerline{ \it $^a$Harish-Chandra Research Institute, 
Chhatnag Road, Jhusi,
Allahabad 211019, India}

\centerline{ \it $^b$School of Physics, 
Korea Institute for Advanced Study, Seoul 130-722, Korea}

\centerline{ \it $^c$International Centre for Theoretical Sciences,
Malleshwaram,
Bengaluru - 560 012, India.}

\vspace*{1.0ex}
\centerline{\small E-mail:  
sitenderpratap,swapno,sen,mritunjayverma@mri.ernet.in}

\vspace*{5.0ex}

\centerline{\bf Abstract} \bigskip

In a metastable de Sitter space any object has a finite 
life expectancy beyond which it undergoes vacuum decay. 
However, by spreading into different parts of the
universe which will fall out of causal contact of each other in future,
a civilization
can increase its collective life expectancy, defined as the average time
after which the last settlement disappears due to vacuum decay. 
We study in detail the collective life expectancy of two comoving 
objects in de Sitter space as a function of the initial separation,
the horizon radius and the vacuum decay rate. We find that
even with a modest initial separation, the collective life expectancy
can reach a value close to the maximum possible value of 1.5 times that
of the individual object if the decay rate is less than 1\% of the
expansion rate. Our analysis can be generalized to any number of objects,
general trajectories not necessarily at rest in the comoving coordinates 
and general FRW space-time. As  part 
of our analysis we find that
in the current state of the universe dominated by matter and cosmological
constant, the vacuum decay rate is increasing as a function of time due to
accelerated expansion of the volume of the past light cone. 
Present decay
rate is about 3.7 times larger than the average decay rate in the 
past and
the final decay rate in the cosmological constant dominated epoch will
be about 56 times larger than the average decay 
rate in the past.
This considerably
weakens the lower bound on the half-life of our universe based on its 
current age.

\vfill \eject

\baselineskip=18pt

\tableofcontents

\sectiono{Introduction} \label{sintro}

The possibility that we may be living in a metastable vacuum has been
explored for more that fifty 
years\cite{okun,stone,frampton,coleman1,coleman2,coleman}.
Discovery of the accelerated expansion of the 
universe\cite{9805201,9812133} 
and subsequent
developments in string theory leading to the construction of
de Sitter vacua\cite{0004134,0105097,0301240} 
suggest that the vacuum we are living in at present is indeed
metastable. Unfortunately our understanding of string theory has
not reached a  stage where we can make a definite prediction about the
decay rate of our vacuum. 
The only information we have about this is
from the indirect observation that our universe is about 
$1.38\times 10^{10}$
years old. Therefore, assuming that we have not been extremely lucky
we can conclude that 
our inverse decay rate\footnote{For
exponential decay the inverse
decay rate differs from half-life by a factor of $\ln 2$. In order 
to simplify
terminology, we shall from now on use only
inverse decay rate and
life expectancy -- to be defined later -- as measures of longevity.} is at least of the \changek{same order}.\footnote{We shall in fact see in \S\ref{smatter} that 
the actual
lower bound for the current inverse decay rate is weaker by a factor of 3.7, 
making it
comparable to the time over which the
earth will be destroyed due to the increase in the size of the sun.
Allowing for the possibility that we could have been extremely lucky
reduces the lower bound on the
inverse decay rate by about a factor of 10\cite{bostron,0512204}.}

Typically the decay of a metastable vacuum proceeds via bubble 
nucleation\cite{okun,stone,frampton,coleman1,coleman2,coleman}
\changed{(see \cite{1505.06397} for a recent survey)}. 
In a small region of space-time the universe 
makes transition to a more stable vacuum, and this bubble of
stable vacuum\footnote{We shall refer to the more stable vacuum as the
stable vacuum, even if this vacuum in turn could decay to other vacua
of lower energy density. In any case since this vacuum will have negative
cosmological constant, the space-time inside the bubble will undergo a
gravitational crunch\cite{coleman}. We shall ignore the possibility of
decay to Minkowski vacua or other de Sitter vacua of lower cosmological
constant since the associated decay rates are very small due to smallness
of the cosmological constant of our vacuum.} 
then expands \changed{at a speed that asymptotically approaches the
speed of light,} converting the rest 
of the region it encounters also to this stable phase. 
\changed{Due to 
this rapid expansion rate it is impossible to observe the expanding bubble
before encountering it -- it reaches us when we see it.}
However, due to the
existence of the future horizon in the de Sitter space, 
even a bubble expanding at the speed
of light cannot fill the whole space at future infinity. Indeed, 
it has been known for quite
some time that in de Sitter space if the expansion rate of the universe 
exceeds the decay rate due to phase transition then even collectively
the bubbles of
stable vacuum cannot fill the whole space\cite{guth} and there will
always be regions which will continue to exist in the 
metastable vacuum.  Nevertheless, any single observer in the metastable
vacuum will sooner or later encounter an expanding bubble of stable
vacuum, and the probability of this decay per unit time determines the
\changem{inverse decay rate} of the observer in the metastable vacuum.

 This suggests that while any single 
 observer will always have a \changem{limited average life span
 determined by the microscopic physics}, a civilization could
 collectively increase its \changem{longevity} by spreading out 
 and establishing different civilizations in different 
 parts of the universe\cite{1503.08130}.
 If the bubble of stable vacuum hits the civilization -- henceforth refered
 to as object -- in the initial stages 
 of spreading out then it does not help since the same bubble will
 most likely destroy all the objects. However, with time the different 
 objects will go outside each other's horizon and a single bubble
 of stable vacuum will not be able to destroy all of them. This will clearly
 increase the life expectancy of the objects collectively -- defined
 as the average value of the time at which the last 
 \changes{surviving object} undergoes vacuum
 decay --
 although there will be no way of telling {\it a priori} which one will 
 survive the longest.
 A simple calculation shows that if we could begin with 2 \changes{objects} 
 already far outside each other's horizon so that their 
 decay probabilities can be taken to be independent, then the life expectancy
 of the combined system
increases by a factor of 3/2 compared to the life expectancy of a single isolated
 object. In the case of $n$ copies the \changen{life expectancy} 
 increases by a factor 
 given
 by the $n$-th harmonic number. However, in actual practice 
 we \changes{cannot 
begin} with copies of the object already outside each other's
 horizon. As a result 
 the increase in the life expectancy is expected to be lower.

\begin{figure}
\begin{center}
\epsfbox{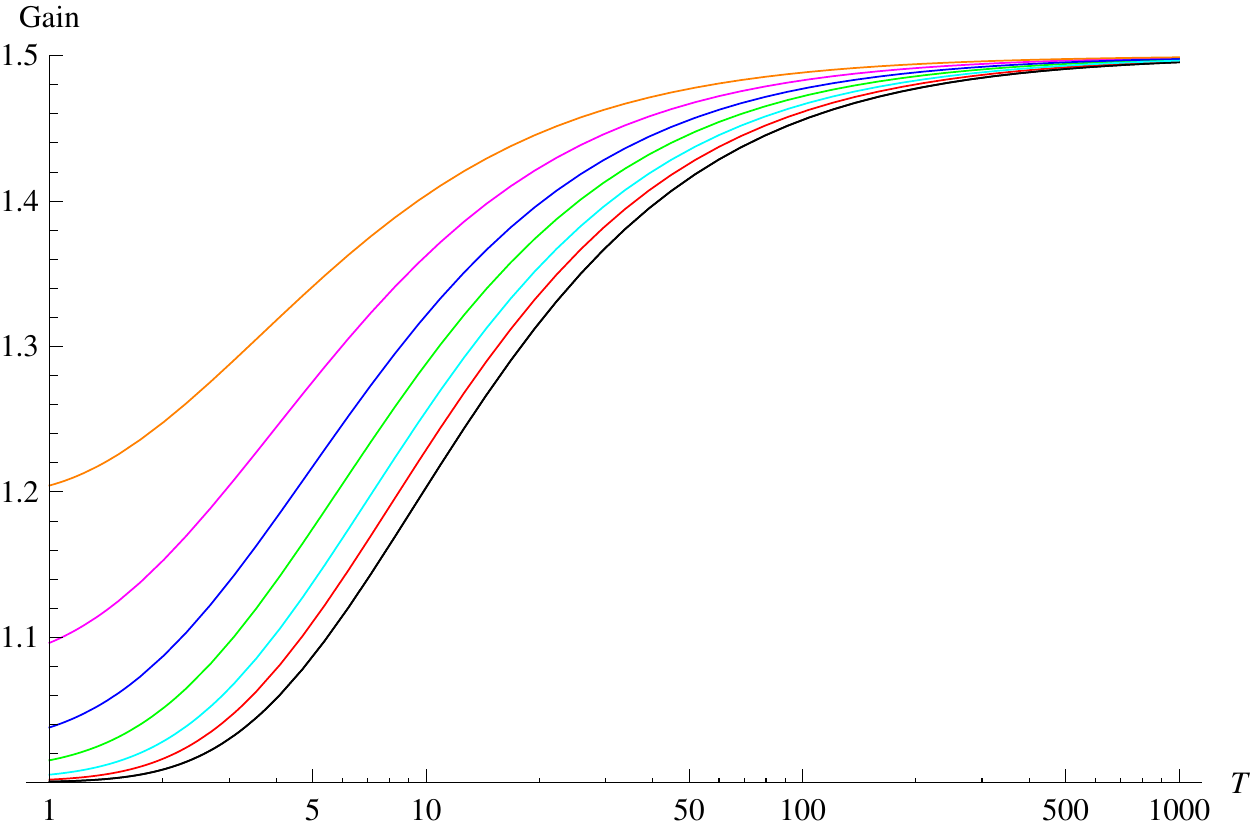}
\caption{The figure showing the \changes{`gain'} in the life expectancy for two
objects compared to that of one object as a function of $T$ for
$r=$ .0003, \tcr{.001}, \tccy{.003}, \tcg{.01}, \tcb{.03}, \tcv{.1}
and \tcy{.3}. \label{fnum}}
\end{center}
\end{figure}

The goal of this paper will be to develop a systematic procedure for computing
the increase in the life expectancy of the object as a result of making 
multiple copies of itself. For two objects we obtain explicit expression
for the life expectancy
in terms of three parameters: the Hubble constant $H$ of the de Sitter
space-time determined by the cosmological constant, the vacuum decay rate
or equivalently the life expectancy $T$ of a single isolated object and the
initial separation $r$ between the two  objects. In fact due to dimensional
reasons the  result depends only on the combination $HT$ and $Hr$, so we
work by setting $H=1$. In Fig.~\ref{fnum} we have shown 
the result for the ratio of the life expectancy of two
objects to that of a single object -- called the \changes{`gain'} --
as a function of $T$ for different
choices of $r$. \changek{From this we see that even for a modest value of 
$r=3\times 10^{-4}$ the gain in the life expectancy reaches close
to the maximum possible value of 1.5 if $T$ is larger that 100 times the
horizon size of the de Sitter space, i.e. the decay rate is less than
$1\%$ of the expansion rate. $T=100$ corresponds to about $1.7\times
10^{12}$  years. 
$r=3\times 10^{-4}$ corresponds to a physical distance of the order of 
$5\times 10^6$ light years
and is of the order of the minimal distance needed to escape the local
gravitationally bound system of galaxies.}   
If $T=10$ -- i.e. of order \changek{$1.7\times 10^{11}$} years --
the gain is about 20\% \changek{for $r=3\times 10^{-4}$}. These \changek{time} 
scales are
\changes{shorter than the 
time scale by which all the stars in the galaxy 
will die. 
Therefore, if $T$ lies between
$10^{11}$ years and the life span of the last star in the local group of
galaxies which will be gravitationally bound and will remain inside 
each other's horizon, 
then we gain \changed{a factor of  1.2 - 1.5} in life expectancy even
by making one additional copy of the object at a distance larger than
about $10^7$ light years from us. On the other hand if 
$T$ is larger than the life span of the
last star in the
galaxy then our priority should be to plan how to survive the death of the galaxy
rather than vacuum decay.} Some discussion on this can be found in
\cite{9902189}.

\begin{figure}
\begin{center}
\epsfbox{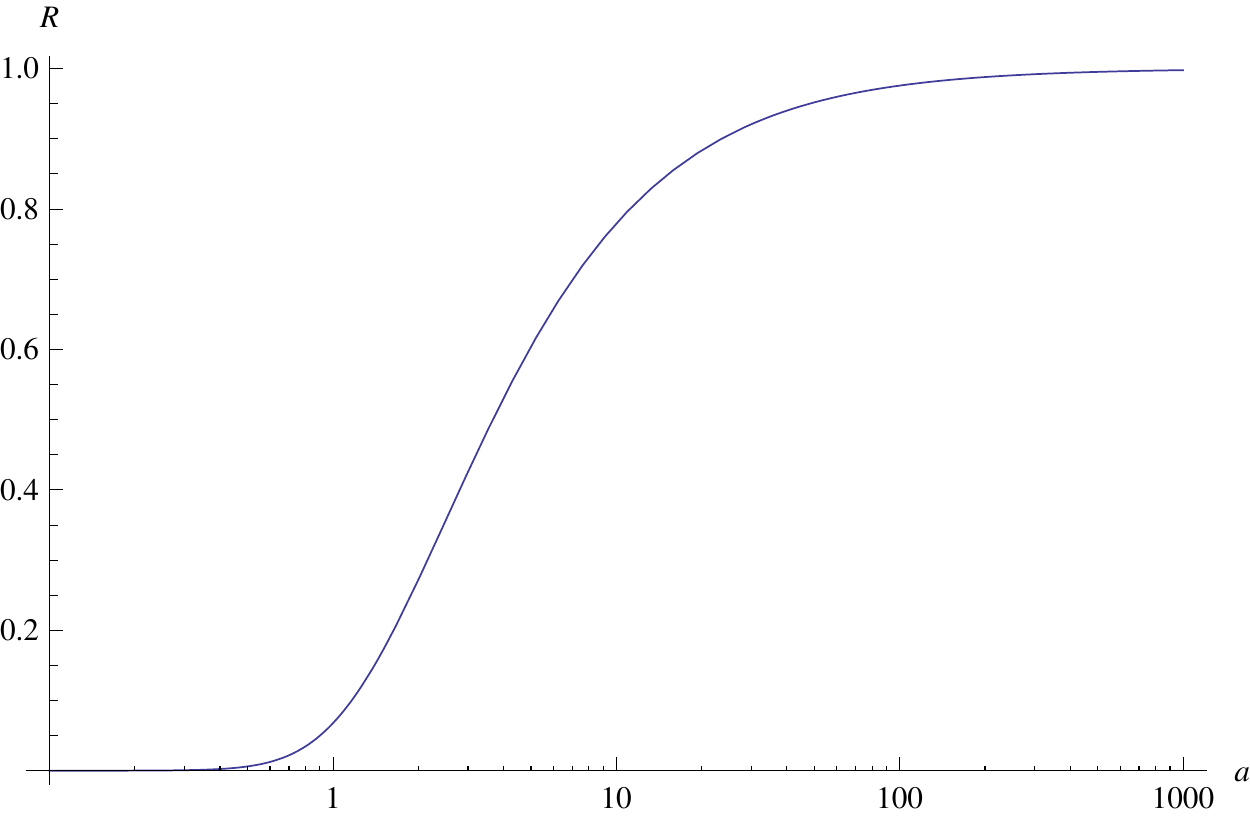}
\end{center}
\caption{\changem{Growth of the relative decay rate $R$ -- defined as the
ratio of the decay rate to its asymptotic value --
of a single
object  in FRW space-time
as a function of the scale factor $a$. Value of $R$ at
$a=1$ represents the decay
rate today relative to what it would be in the cosmological constant
dominated epoch. \label{frate}}}
\end{figure}

Even though most of our analysis focusses on the case of a pair of objects 
in de Sitter space at fixed comoving coordinates, our method is quite 
general and can be applied to arbitrary number of objects 
in a general
FRW metric moving along general trajectories. We \changes{discuss these}
generalizations in \S\ref{sgen}. In particular considering the case of a single
object in an FRW metric dominated by matter and cosmological constant,
as is the case with the current state of our universe, we find that the vacuum
decay rate increases as a function of time due to accelerated expansion of
the volume of the past light cone. 
\changem{This has been shown in
Fig.~\ref{frate}}.
This rate approaches a constant value
as the universe enters the cosmological constant dominated era, but we find
for example that this asymptotic decay rate is about \changed{15} 
times larger than
the decay rate today, \changem{which in turn is about $3.7$
times larger than the average  decay
rate in the past}.
\changek{Now given that 
the universe has survived for
about $1.38\times 
10^{10}$ years, we can put a lower bound of this order}
on the inverse of the average decay rate in the 
past.\footnote{\changem{One must keep in mind that this is not a 
strict bound since we could have survived till today by just being lucky.}}
\changem{This translates to a lower bound of order
$3.7\times 10^9$ years on the \changem{inverse decay rate} today and
$2.5\times 10^8$ years on the asymptotic
\changem{inverse decay rate}.}

The rest of the paper is organised as follows. In \S\ref{sindependent}
we describe the case of the decay of $n$ objects assuming
that their decay probabilities are independent of each other, and show that
the life expectancy of the combined system goes up by a 
factor \changes{equal to} the $n$-th
harmonic number. In \S\ref{s11} we carry out the complete analysis for two
observers in 1+1 dimensional de Sitter space.
The final result for the life expectancy of the combined system can be 
found in \refb{ebt12fin}. This is generalized to the case
of two observers in 3+1 dimensional de Sitter space-time in \S\ref{s31}.
Eq.\refb{ep12four} 
together with \refb{epp12} and \refb{3.1.24} gives the \changes{probability that
at least one of the two objects survives till time $t$}, which can then be used
to compute the life expectancy of the combined system using \refb{et12four}.
In \S\ref{sgen} we discuss various generalizations including the case of multiple
observers, general trajectories and general FRW type metric. 
We conclude in \S\ref{sdiss} with a discussion of how in future
we could improve our knowledge of
possible values of the parameters $r$ and $T$ which enter our calculation.
In appendix \ref{sapp} we compute the time dependence of the decay rate
for a general equation of state of the form $p=w\rho$.

\sectiono{Independent decay} \label{sindependent}

Let us suppose that we have two independent objects, each with a decay
rate of $c$ per unit time. We shall label them as $C_1$ and $C_2$.
If we begin \changed{with the assumption that both objects exist}
at time $t=0$ then the probability that
the first object exists after time $t$ is
\be 
P_1(t) = e^{-c\, t}\, .
\ee
Therefore, the probability that it decays between time $t$ and $t+\delta t$ is
$\changen{- \dot P_1(t)\delta t}$ \changek{where $\dot P_1$
denotes the derivative of $P_1$ with respect to $t$}, and its life expectancy, is
\be
\bar t_1 = -\int_0^\infty t \changen{\dot P_1(t)} dt = \int_0^\infty P_1(t) dt = c^{-1}\, .
\ee
Independently of this the probability that the second object exists after time
$t$ is
also given by $\exp[-c\, t]$ 
and it has the same life expectancy.

Now let us compute the life expectancy of both objects combined, defined as
the average of the larger of the actual life time of $C_1$ and $C_2$.
To compute this note that since the two objects are independent,
the probability that both will decay by time $t$ is
given by $(1-P_1(t)) (1-P_2(t))=(1-P_1(t))^2$. Therefore, the probability that
the last one to survive decays between $t$ and $t+\delta t$ is 
${d\over dt} (1-P_1(t))^2 \delta t$. This gives the life expectancy of the
combined system to be
\be \label{eindep}
\bar t_{12}= \int_0^\infty t\, {d\over dt}  (1-P_1(t))^2 dt = {3\over 2} c^{-1}\, .
\ee
Therefore, we see that by taking two independent objects we can increase the
life expectancy by a factor of $3/2$. A similar argument shows that for $n$
independent objects the life expectancy will be
\be 
\bar t_{12\cdots n}=
\int_0^\infty t\, {d\over dt}  (1-P_1(t))^n dt = \left(1+{1\over 2} +{1\over 3} +\cdots
+{1\over n}\right) c^{-1}\, .
\ee

\sectiono{Vacuum decay in 1+1 dimensional de Sitter space} \label{s11}

Consider 1+1 dimensional de Sitter space
\be
ds^2 = -dt^2 + e^{2t} dx^2 \, .
\ee
Note that we have set the Hubble constant of the de Sitter space 
and the speed of light to unity so that
all other time / lengths appearing in the analysis are to be interpreted as their
values in units of the inverse Hubble constant.  We introduce the conformal
time $\uptau$ via
\be
\uptau = -e^{-t}
\ee
in terms of which the metric takes the form
\be
ds^2 = \uptau^{-2} (-d\uptau^2 + dx^2)\, .
\ee
At $t=0$ we have $\uptau=-1$ and comoving distances coincide with the
physical distances.

\changek{We shall use this space-time as a toy model for
studying the kinematics of vacuum decay. We shall assume that in this
space-time there is a certain probability per unit time per unit volume 
of producing a bubble of stable
vacuum, which then expands at the speed of light causing decay of the
metastable vacuum. 
We shall not
explore how such a bubble is produced; instead our goal will be
to study its effect on the life expectancy of
the objects living in this space. In \S\ref{s31} we shall generalize this
analysis to 3+1 dimensional de Sitter space.}

\begin{figure}
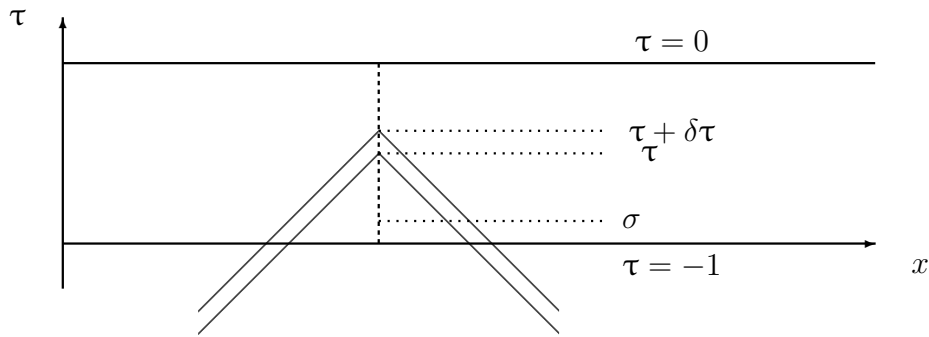

\begin{center}
\figtwo
\end{center}
\vskip -.5in
\caption{A comoving object in de Sitter space and
\changen{its}  past light cones at
conformal times $\uptau$ and $\uptau+\delta\uptau$.
\label{f2}}
\end{figure}

\subsection{Isolated comoving object} \label{sisolated}

Consider a single object in de Sitter space at rest in the comoving coordinate
$x$ (say at $x=0$), shown by the vertical dashed line in Fig.~\ref{f2}.  
We start at $t=0$ ($\uptau=-1$) and are interested in calculating the 
probability
that it survives at least till \changec{conformal}
time $\uptau$. If we denote this by $P_0(\uptau)$ then the
probability that it will decay between $\uptau$ and $\uptau+\delta\uptau$ is
$-P_0'(\uptau) \delta\uptau$ 
\changek{where $'$ denotes derivative with respect to $\uptau$}. 
On the other hand this probability is also 
\changed{given} by the
product of $P_0(\uptau)$ and the probability that a vacuum bubble is produced
somewhere in the past light cone of the object between $\uptau$ and 
$\uptau+\delta\uptau$, as shown in Fig.\ref{f2}. The volume of the past light cone
of this interval can be easily calculated to be
\be \label{epast}
2 \, \delta\uptau \int_{-\infty}^\uptau {d\sig\over \sig^2} = -{2\over \uptau}\, \delta\uptau\, .
\ee
Therefore, if $K$ is the probability of producing the bubble per unit space-time
volume 
then the probability of producing a bubble in the past light cone of the
object between $\uptau$ and $\uptau+\delta\uptau$ is given by
$-2K \delta\uptau/\uptau$. The previous argument then leads to the equation
\be \label{epast2}
P_0'(\uptau) =  2 K \uptau^{-1} P_0(\uptau) \, .
\ee
This equation, together with the boundary condition $P_0(\uptau=-1)=1$,
can be integrated to give
\be
\ln P_0(\uptau) = 2 K \ln (-\uptau)\, .
\ee
In terms of physical time $t$ we have\footnote{\changem{Note that by an abuse of 
notation we have used the same symbol $P_0$ to denote
the probability as a function of $t$ although the functional form changes. We 
shall continue to follow this convention, distinguishing the function by its
argument ($t$ or $\uptau$)}. \changen{Derivatives with respect to $\uptau$ and $t$ will 
be distinguished by using $P_0'$ to denote $\uptau$-derivative of $P_0$ and
$\dot P_0$ to denote $t$-derivative of $P_0$.}}
\be
P_0(t) = e^{-2 K t}\, .
\ee
From this we can calculate the life expectancy, defined as the
integral of $t$ weighted by the probability that the object undergoes vacuum
decay between $t$ and $t+dt$. Since the latter is given by 
$\changen{-\dot P_0(t)} dt$, we have
the life expectancy
\be \label{et0}
T = -\int_0^\infty t\, \changen{\dot P_0(t)} 
\, dt = \int_0^\infty  P_0(t) \, dt = {1\over 2K}\, ,
\ee 
where in the second step we have used integration by parts.
We shall express our final results in terms of $T$ instead of
$K$.

\begin{figure}
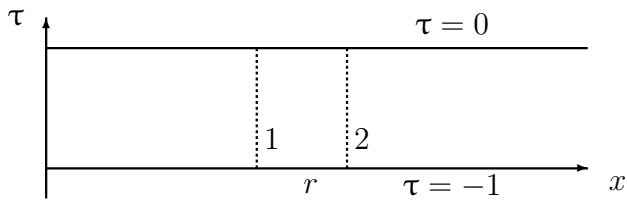

\begin{center}
\figone
\end{center}
\vskip -.5in
\caption{Two comoving objects in de Sitter space separated by
physical distance $r$ at $\uptau=-1$.
\label{f1}}
\end{figure}

\subsection{A pair of comoving objects}

Next we shall consider two comoving objects $C_1$ and $C_2$
in de Sitter space
separated by physical distance $r$ at $t=0$ or equivalently $\uptau=-1$.
We shall take $r<1$, i.e.\ assume that the two objects are within each other's
horizon at the time they are created. 
We denote by $P_i(\uptau)$ the probability that $C_i$ survives at
least till conformal time $\uptau$ for $i=1,2$ and by
$P_{12}(\uptau_1,\uptau_2)$ the joint probability that $C_1$ survives at
least till conformal time $\uptau_1$ and $C_2$ survives at least
till conformal time $\uptau_2$. The boundary condition will be set by assuming
that both objects exist \changec{at $\uptau=-1$}, so that we have 
\be \label{eboundary}
P_1(-1)=1,
\quad P_2(-1)=1, \quad P_{12}(-1,\uptau_2)=P_2(\uptau_2),  \quad 
P_{12}(\uptau_1,-1) = P_1(\uptau_1) \, .
\ee

\begin{figure}
\begin{center}
\figthree
\end{center}
\vskip -.5in
\caption{The past light-come of $C_2$ at $\uptau=-1$ and the past 
light cone of $C_1$ between $\uptau$ and $\uptau+\delta\uptau$ for
$\uptau<r-1$.
\label{f3}}
\end{figure}

First we shall calculate $P_1(\uptau)$ and $P_2(\uptau)$. They must be
identical by symmetry, so let us focus on $P_1(\uptau)$.  The calculation
is similar to that for $P_0(\uptau)$ above for a single isolated object, except
that the existence of $C_2$ at $\uptau=-1$ guarantees that no vacuum
decay bubble is produced in the past light-come of $C_2$ at $\uptau=-1$, 
and hence
while computing the volume of the past light cone of the $C_1$ between
$\uptau$ and $\uptau+\delta\uptau$, we have to exclude the region inside
the past light cone of $C_2$ at $\uptau=-1$. This has been shown in 
Fig.~\ref{f3}. This volume is given by
\be
\delta\uptau\left[2\int_{-\infty}^\uptau {d\sig\over \sig^2} 
- \int_{-\infty}^{-1 - {r-1-\uptau\over 2}}
{d\sig\over \sig^2} \right] = \delta\uptau \left[-{2\over \uptau} 
- {2\over r+1-\uptau}\right]
\quad \hbox{for $\uptau< r-1$}\, .
\ee
However, for $\uptau>r-1$ the past light cone of $C_1$ between $\uptau$ and
$\uptau+\delta\uptau$ does not intersect the past light cone of $C_2$ at
\changed{$\uptau=-1$} (see Fig.~\ref{f4}), and we get the volume to be
\be
2\, \delta\uptau\, \int_{-\infty}^\uptau {d\sig\over \sig^2}  = - 2 \, \delta\uptau \, 
{1\over \uptau}
\quad \hbox{for $\uptau> r-1$}\, .
\ee

\begin{figure}
\begin{center}
\figfour
\end{center}
\vskip -.5in
\caption{The past light-come of $C_2$ at $\uptau=-1$ and the past 
light cone of $C_1$ between $\uptau$ and $\uptau+\delta\uptau$ for $\uptau>r-1$.
\label{f4}}
\end{figure}

This leads to the following differential equation for $P_1(\uptau)$:
\be \label{e3.12}
{1\over P_1(\uptau)}{dP_1\over d\uptau} = 
%\begin{cases}
\begin{cases}
{- K \left[-{2/\uptau} - {2/ (r+1-\uptau)}\right]
\quad \hbox{for $\uptau< r-1$}\, ,}  \cr
{2\, K\,   /\uptau
\quad \hbox{for $\uptau> r-1$}\, . }
\end{cases}
\ee
Using the boundary condition $P_1(-1)=1$ and the continuity of $P_1(\uptau)$
across $\uptau=r-1$ we get
\be \label{ep1}
\ln P_1(\uptau) = \begin{cases} 2 \, K\, \{ \ln (-\uptau) - \ln(r+1-\uptau) + \ln (r+2)\} \quad
\hbox{for $\uptau< r-1$}\, , \cr
2\, K \, \{ \ln(-\uptau) - \ln 2 + \ln (r+2)\} \quad \hbox{for $\uptau> r-1$}\, .
\end{cases}
\ee
Using the symmetry between 1 and 2 we also get the same expression for
$P_2(\uptau)$.
In terms of the physical time $t$ we have
\be \label{epitexp}
P_1(t)=P_2(t)= \begin{cases}
e^{-2Kt} (r+1+e^{-t})^{-2K} (r+2)^{2K} 
\quad \hbox{for $t< -\ln(1-r)$}\, , \cr
\Big((r+2) / 2\Big)^{2K} e^{-2Kt} \quad \hbox{for $t> -\ln(1-r)$}\, .
\end{cases}
\ee
Therefore, the life expectancy of  $C_1$ is
\ben \label{et1}
\bar t_1 &=& -\int_0^\infty t\, \dot P_1(t) \, dt =\int_0^\infty P_1(t) dt
\nonumber \\ &=&
(r+2)^{2K} \left[B\left({1\over 2+r}; 2K, 0\right) - B\left( {1-r\over 2};2K,0
\right) + {(1-r)^{2K}\over 2^{2K+1} K}\right]
\een
where $B(x;p,q)$ is the incomplete beta function, defined as
\be
B(x;p,q) =  \int_0^x t^{p-1} (1-t)^{q-1} dt = 
\int_0^{x/(1-x)} {y^{p-1}\over (1+y)^{p+q}} dy\, ,
\ee
the two expressions being related by the transformation $t=y/(y+1)$.
In terms of the life expectancy $T=1/2K$ of a single isolated object,
we have
\be \label{eavt1}
\bar t_1 = (r+2)^{1/T} \left[B\left({1\over 2+r}; {1\over T},0\right) 
- B\left( {1-r\over 2}; {1\over T},0
\right) + T \, {(1-r)^{1/T}\over 2^{1/T} }\right]\, .
\ee
$C_2$ also has the same life expectancy. \refb{eavt1}
is somewhat larger \changen{than $T$}, but
that is simply a result of our initial assumption that both objects exist
at $t=0$. 
If both objects had started at the same space-time point and then got
separated following some specific trajectories, then there would have been
a certain probability that one or both of them will decay during the process
of separation; this possibility has been ignored here leading to the apparent
increase in the life expectancy. However, for realistic 
values of $r$
and $T$, which corresponds to $r<<1$ and $T\gsim 1$, the ratio 
$\bar t_1/T$ remains close to unity.

Let us now turn to the \changed{computation of the  joint 
survival probability $P_{12}(\uptau_1,\uptau_2)$. In this}
case the probability that the first object undergoes vacuum decay between
$\uptau_1$ and $\uptau_1+\delta\uptau_1$ and the second object survives at
least till $\uptau_2$ is given by $-\delta\uptau_1 \left(\p P_{12}(\uptau_1,\uptau_2)/\p\uptau_1
\right)$. On the other hand the same probability is given by 
$K\times P_{12}(\uptau_1,\uptau_2)$ times the
volume of the past light-come of $C_1$ between $\uptau_1$ and $\uptau_1+
\delta\uptau_1$, excluding the region inside the past light cone of 
\changed{$C_2$} 
at $\uptau_2$. The relevant geometry has been shown in Figs.~\ref{f5},
\ref{f6} and \ref{f7} for different ranges of $\uptau_1$ and $\uptau_2$. The
results are as follows:
\begin{enumerate}
\item For $\uptau_1<\uptau_2-r$ the geometry is shown in Fig.~\ref{f5}. In this case
\changec{$C_1$ at $\uptau_1$ (and hence} 
the whole of the past light cone of \changec{$C_1$} between $\uptau_1$ and
$\uptau_1+\delta\uptau_1$) is inside the past light cone of $C_2$ at $\uptau_2$.
Therefore, the decay probability is zero and we have the equation:
\be \label{efirst}
{\p \ln P_{12}(\uptau_1,\uptau_2)\over \p\uptau_1} = 0 \quad \hbox{for $\uptau_1<\uptau_2-r$}\, .
\ee

\begin{figure}
\begin{center}
\figfive
\end{center}
\vskip -.3in
\caption{The past light-come of $C_2$ at $\uptau_2$ and the past 
light cone of $C_1$ between $\uptau_1$ and $\uptau_1+\delta\uptau_1$ for 
$\uptau_1<\uptau_2-r$.
\label{f5}}
\end{figure}

\item For $\uptau_2-r<\uptau_1<\uptau_2+r$ the geometry is as shown in Fig.~\ref{f6}.
In this case \changec{$C_1$ at $\uptau_1$ and $C_2$ at $\uptau_2$ 
are space-like separated. 
The volume} of the past light cone \changed{of $C_1$} between
$\uptau_1$ and $\uptau_1+\delta\uptau_1$ outside the past light cone of 
\changed{$C_2$} at $\uptau_2$ is given by
\be
\int_{-\infty}^{\uptau_1} {d\sig\over \sig^2} 
+ \int_{{1\over 2}(\uptau_1+\uptau_2-r)}^{\uptau_1} 
{d\sig\over \sig^2} 
= -{2\over \uptau_1} - {2\over r-\uptau_1-\uptau_2}\, .
\ee
This gives
\be \label{esecond}
{\p \ln P_{12}(\uptau_1,\uptau_2)\over \p\uptau_1} = 2 K 
\left\{ {1\over\uptau_1} + {1\over r-\uptau_1-\uptau_2} \right\} \quad 
\hbox{for $\uptau_2-r<\uptau_1<\uptau_2+r$}\, .
\ee

\begin{figure}
\begin{center}
\figsix
\end{center}
\vskip -.4in
\caption{The past light-come of $C_2$ at $\uptau_2$ and the past 
light cone of $C_1$ between $\uptau_1$ and $\uptau_1+\delta\uptau_1$ for 
$\uptau_2-r<\uptau_1<\uptau_2+r$.
\label{f6}}
\end{figure}

\item For $0<\uptau_2+r<\uptau_1$, the geometry is shown in \changec{Fig.~\ref{f7}}. 
In this
case \changec{$C_2$ at $\uptau_2$ is inside the past light cone of $C_1$ 
at $\uptau_1$ and} 
there is no intersection between the past light cone of $C_1$ between
$\uptau_1$ and $\uptau_1+\delta\uptau_1$ and the past light cone of $C_2$ at
$\uptau_2$. Therefore, the volume of the past light cone of $C_1$ between
$\uptau_1$ and $\uptau_1+\delta\uptau_1$ is given by
\be
2 \int_{-\infty}^{\uptau_1} {d\sig\over \sig^2} = -{2\over \uptau_1}\, ,
\ee
\changed{and} we have
\be \label{ethird}
{\p \ln P_{12}(\uptau_1,\uptau_2)\over \p\uptau_1} = 2 K 
 {1\over\uptau_1}
\quad \hbox{for $\uptau_2+r<\uptau_1<0$}\, .
\ee
\begin{figure}
\vskip .3in
\begin{center}
\figseven
\end{center}
\vskip -.4in
\caption{The past light cone of $C_2$ at $\uptau_2$ and the past 
light cone of $C_1$ between $\uptau_1$ and $\uptau_1+\delta\uptau_1$ for 
$\uptau_2+r<\uptau_1<0$.
\label{f7}}
\end{figure}

\end{enumerate}
We can now determine $P_{12}(\uptau_1,\uptau_2)$ by integrating \refb{efirst},
\refb{esecond}, \refb{ethird} subject to the boundary condition given in
\refb{eboundary}
\be \label{eboundary1}
P_{12}(-1,\uptau_2) = P_2(\uptau_2) \, \changed{ = P_1(\uptau_2)}\, ,
\ee
and using the fact that $P_{12}(\uptau_1,\uptau_2)$ must be continuous across the
subspaces defined by $\uptau_1=\uptau_2\pm r$. The result of the integration is
\be
\ln P_{12}(\uptau_1,\uptau_2) = 
 \begin{cases} 2K \left\{ \ln(-\uptau_2) +  \ln (r+2) - \ln 2 \right\}
\quad \hbox{for $\uptau_1<\uptau_2-r$} \, ,\cr
2K \{ \ln(-\uptau_2) +\ln(-\uptau_1) -\ln(r-\uptau_1-\uptau_2) + \ln (r+2)\}\, ,
\quad 
\hbox{for $\uptau_2-r<\uptau_1<\uptau_2+r$}\cr
2K\{ \ln (-\uptau_1) + \ln(r+2) -\ln 2\}\quad \hbox{for $\uptau_2+r<\uptau_1<0$} \, .
\end{cases}
\ee
Note that the result is symmetric under the exchange of $\uptau_1$ and $\uptau_2$
even though at the intermediate stages of the analysis 
\changek{this symmetry was not manifest}.

Expressed in terms of physical time the above solution takes the form:
\be \label{ep12}
P_{12}(t_1,t_2)= 
 \begin{cases} \{(r+2)/2\}^{2K} e^{-2Kt_2}
\quad \hbox{for $t_1<-\ln (r+e^{-t_2})$} \, ,\cr
(r+2)^{2K} e^{-2K(t_1+t_2)} (r+e^{-t_1}+e^{-t_2})^{-2K} 
\quad 
\hbox{for $-\ln (r+e^{-t_2})< t_1 < -\ln(e^{-t_2}-r)$}\, ,\cr
\{(r+2)/2\}^{2K} e^{-2Kt_1} \quad \hbox{for $t_1>-\ln(e^{-t_2}-r)$} \, .
\end{cases}
\ee
\changed{If $e^{-t_2}-r$} is negative then the third case is not relevant and in
the second case there will be no upper bound on $t_1$. \changec{Physically 
this can be understood by noting that in this case $\uptau_2> -r$ and $C_2$
will never come inside the past light cone of $C_1$ even when $\uptau_1$
reaches its maximum value 0.}

\begin{figure}
\begin{center}
 \figvenn
 \end{center}
 \caption{Probability rule  for N=2 using Venn Diagram.}	
 \label{fig.1}
\end{figure}

Our interest lies in computing the probability that at least
one of the two objects survives till time $t$.
Let us
denote this by $\wt P_{12}(t)$. This is given by the sum of the probability
that \changec{$C_1$ survives till time $t$ and the probability that
$C_2$ survives till time $t$, but we have to subtract from it the
probability that both $C_1$ and $C_2$} 
survive till time $t$ since this will be counted
twice otherwise.  This can be seen from the Venn diagram of two objects
shown in Fig.~\ref{fig.1}. Therefore, we have
\be \label{ewtp12}
\wt P_{12}(t) = P_1(t)+P_2(t) - P_{12}(t,t)\, .
\ee
From this we can compute the probability that the last one to survive 
decays between $t$ and
$t+\delta t$ as
\be 
 -\delta t\, {d\over dt} \wt P_{12}(t)\, .
\ee
Therefore, the life expectancy of the combined system is given by
\be \label{eavlife}
\bar t_{12} = -\int_0^\infty dt \, t\, {d\over dt} \wt P_{12}(t,t)
=\int_0^\infty dt \,  \{P_1(t) + P_2(t) - P_{12}(t,t)\}\, ,
\ee
where in the second step we have integrated 
by parts and used \refb{ewtp12}.
\changed{Each of the first two integrals gives} 
the result $\bar t_1$ computed in
\refb{eavt1}. For the last integral since we have to evaluate $P_{12}(t_1,t_2)$
at $t_1=t_2=t$ only the middle expression in \refb{ep12} is relevant, and we get
\ben 
\int_0^\infty P_{12}(t,t) dt &=& 
\int_0^\infty (r+2)^{2K} e^{-4Kt} (r+2 e^{-t})^{-2K}  dt \nonumber \\
&=& 2^{-4K} r^{2K} (r+2)^{2K} B\left( {2\over 2+r}; 4K, -2K\right)\, .
\een
Combining this with the result for $\bar t_1$ given in \refb{eavt1}
and replacing $K$ by $1/2T$ we get
\ben \label{ebt12fin}
\bar t_{12} &=& 2 (r+2)^{1/T} \left[B\left({1\over 2+r}; {1\over T},0\right) 
- B\left( {1-r\over 2}; {1\over T},0
\right) + T \, {(1-r)^{1/T}\over 2^{1/T} }\right]
\nonumber \\
&& - 2^{-2/T} r^{1/T} (r+2)^{1/T} B\left( {2\over 2+r}; {2\over T}, -{1\over T}
\right)\, .
\een

We can now check various limits. First of \changed{all} we can study the 
$r\to 0$ limit using the result
\be 
B(x;2\alpha, -\alpha) \simeq {1\over \alpha} \, (1-x)^{-\alpha}\, ,
\ee
for $x$ close to 1. This gives $\lim_{r\to 0} \bar t_{12} = T$. This is in agreement
with the fact that if the two objects remain at the same point then their
combined life expectancy is the same as that of individual objects.

If on the other hand we take the limit of large $T$ then, using the result
\be 
B(x;\alpha, \beta) \simeq {1\over \alpha}
\ee
for small $\alpha$, we get
$\bar t_{12} \simeq 3T/2$. Therefore, the life expectancy of the two objects 
together is 3/2 times that of an isolated object. This is consistent
with the fact that if the \changem{inverse decay rate} of individual objects is large then typically
there will be enough time for the two objects to go out of each other's
horizon before they decay. Therefore, we can treat them as independent objects
and recover the result \refb{eindep}. \changec{Mathematically 
this can be seen from the
fact that when $T$ is large and $t\sim T$ then 
$P_{12}(t,t)$ given in the middle expression
of \refb{ep12} approaches  
$e^{-4 K t}$, which in turn is approximately equal to the square of
$P_1(t)$ given in \refb{epitexp}.}

\sectiono{Vacuum decay in 3+1 dimensional de Sitter space} \label{s31}

In this section we shall repeat the analysis of \S\ref{s11} for 3+1 dimensional
de Sitter space-time. Since the logical steps remain identical, we shall
point out the essential differences arising in the two cases and then describe
the results.

The metric of the 3+1 dimensional de Sitter space is given by
\be \label{eflat}
ds^2 = - dt^2 + e^{2t} (dx^2 + dy^2 + dz^2) = \uptau^{-2} (-d\uptau^2 +
dx^2 + dy^2 + dz^2), \quad \uptau\equiv - e^{-t}\, .
\ee
There are of course various other coordinate systems in which we can describe
the de Sitter metric, but the coordinate system used in \refb{eflat} is specially
suited for describing out universe, with $(x,y,z)$ labelling comoving coordinates
and $t$ denoting the cosmic time in which the constant $t$ slices have uniform
microwave background temperature. This form of the metric uses the observed
flatness of the universe. The actual metric at present is deformed due to the
presence of matter density, and also there is a lower cut-off 
\changek{on} $t$ since our
universe has a finite age of the order of the inverse Hubble constant. But both
these effects will become irrelevant within a few Hubble time and we ignore
them. In \S\ref{smatter} we shall study these effects, 
but at present our goal is to get an analytic result under these
simplifying assumptions.

\subsection{Isolated comoving object}

First consider the case of an isolated object.  The calculation proceeds as
in \S\ref{sisolated}. However, in computing the volume of the past light cone
in Fig.~\ref{f2} we have to take into account the fact that for each $\sig$, the
light cone is a sphere of radius $(\uptau-\sig)$.  Since the coordinate
radius of the sphere is $(\uptau-\sig)$ and the space-time volume element 
scales as \changen{$1/\sig^4$ we} get the volume of the past light cone
of the object between $\uptau$ and $\uptau+\delta\uptau$ to be
\be \label{e4.2}
\delta\uptau \int_{-\infty}^\uptau {d\sig\over \sig^4} 4\pi (\uptau-\sig)^2
= -{4\over 3} \pi \uptau^{-1} \delta\uptau\, .
\ee
This replaces the right hand side of \refb{epast}. Therefore, \refb{epast2} 
takes the form
\be \label{e4.3}
P_0'(\uptau) = {4\over 3} \pi \uptau^{-1} K\, P_0(\uptau)\, ,
\ee
with the solution
\be
\ln P_0(\uptau)={4\over 3} \pi K \ln(-\uptau)\, ,
\ee
\be 
P_0 (t) = \exp\left(-{4\over 3}\pi K t\right)\, .
\ee
{}From this we can calculate the life expectancy of the isolated object
to be
\be\label{epav3}
T =\int_0^\infty P_0(t) dt = {3\over 4\pi K}\, .
\ee

\subsection{A pair of comoving objects}

The additional complication in the case of two objects 
comes from having to evaluate the contribution of
the past light come of the first object between $\uptau_1$ and 
$\uptau_1+\delta\uptau_1$ in situations depicted in Figs.~\ref{f3} and
\ref{f6}. Let us consider Fig.~\ref{f6} since Fig.~\ref{f3} can
be considered as a special case of Fig.~\ref{f6} with $\uptau_2=-1$.
Now in Fig.~\ref{f6} which occurs for $\uptau_2-r < \uptau_1<\uptau_2+r$,
the past light cone of $C_1$ between $\uptau_1$ and $\uptau_1+\delta\uptau_1$
lies partly inside the past light cone of $C_2$. We need to subtract
this contribution from the total volume of the past light cone of $C_1$
between $\uptau_1$ and $\uptau_1+\delta\uptau_1$, since the assumption that
$C_2$ survives till $\uptau_2$ rules out the formation of a bubble inside the
past light cone of $C_2$. Our goal will be to calculate this volume.

Examining Fig.~\ref{f6} we see that the intersection of the past light cones
of $C_1$ at $\uptau_1$ and $C_2$ at $\uptau_2$
occur at $\uptau=\sig$ for $\sig<(\uptau_1+\uptau_2-r)/2$. 
At a value of $\sig$ satisfying this constraint, the past light cone of $C_1$
at $\uptau_1$ is a sphere of coordinate
radius $r_1=(\uptau_1-\sig)$ and the past light cone of
$C_2$ at $\uptau_2$ is a sphere of coordinate 
radius $r_2=(\uptau_2-\sig)$. The centers
of these spheres, lying at the comoving coordinates of the two objects
have a coordinate separation of $r$. A simple geometric analysis shows that
the coordinate area of \changed{the}
part of the first sphere that is inside the second
sphere is given by
\be 
\pi {r_1\over r} \, \{ r_2{}^2 - (r_1-r)^2\} = 
\pi {(\uptau_1 - \sig) \over r} (\uptau_2 - \uptau_1 + r) (\uptau_1+\uptau_2 - r - 2 \sig)\, .
\ee
Taking into account the fact that physical volumes are given by $1/\sig^4$ times
the coordinate volume we get the following expression for the volume 
of the past
light cone of $C_1$ between $\uptau_1$ and $\uptau_1+\delta\uptau_1$ that
is inside the past light cone of $C_2$:
\ben \label{excl}
&& {\pi\over r} (\uptau_2 - \uptau_1 + r)  \, \delta\uptau_1 
\int_{-\infty}^{(\uptau_1+\uptau_2-r)/2} {d\sig\over \sig^4}  (\uptau_1 - \sig)
(\uptau_1+\uptau_2 - r - 2 \sig) \nonumber \\ &=& \changes{
\frac{2\, \pi}{3\, r}(\uptau_2-\uptau_1+r)\, \delta\uptau_1\, 
{(3r - \uptau_1 - 3\uptau_2)\over (r - \uptau_1 - \uptau_2)^2}}
\, .
\een
As already mentioned the excluded volume in case of Fig.~\ref{f3} can be
found by setting $\uptau_1=\uptau$ and $\uptau_2=-1$ in \refb{excl}.

\begin{figure}
\begin{center}
\epsfbox{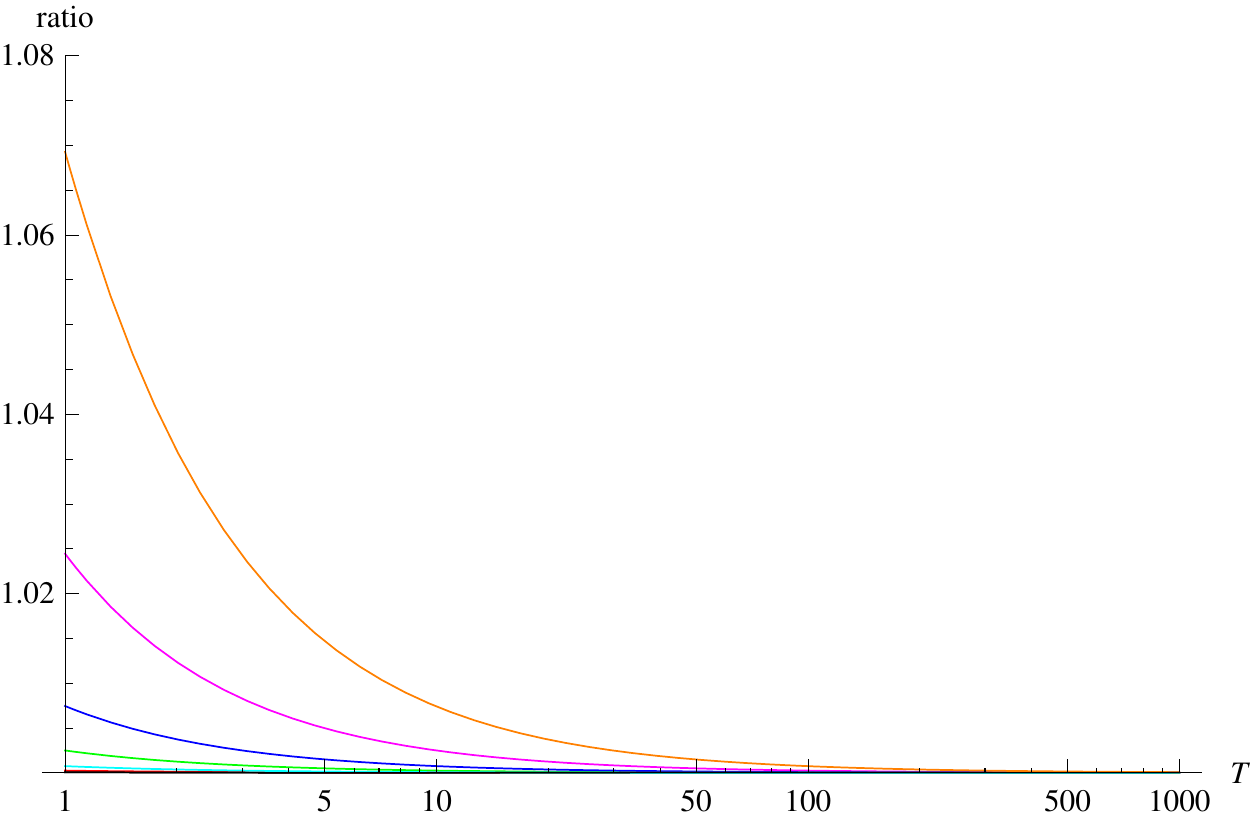}
\caption{The figure showing the ratio $\bar t_1/T$ for
$r=$ .0003, \tcr{.001}, \tccy{.003}, \tcg{.01}, \tcb{.03}, \tcv{.1}
and \tcy{.3}. For $r\le .003$ the ratio is not distinguishable from
1 in this scale. \label{fcomp}}
\end{center}
\end{figure}

We are now ready to generalize all the results of \S\ref{s11}. Let us begin
with \refb{e3.12}. Its generalization to the 3+1 dimensional case takes the form
\be%\[ 
\label{exy1}
\frac{d}{d\uptau} \ln P_1(\uptau)=\left\{
\begin{array}{rl}
& \displaystyle {2\pi K\over 3} \left[ 
\frac{2}{\uptau}+\frac{(-1+r-\uptau)(3+3r-\uptau)}{r(\uptau-r-1)^2}\right]  
\hskip .2in \mbox{if $\uptau<r-1$} \\
&\displaystyle {4\pi K \over 3\uptau} \hskip 2.54in \mbox{if $\uptau>r-1$} 
\end{array} \right.
%\]
\ee
Its solution is given by
\begin{equation}
\ln P_1(\uptau)=\begin{cases}
\frac{4\pi K}{3}\Big(\ln(-\uptau)+\frac{\uptau}{2r}-\ln(-\uptau+r+1)+\frac{2(r+1)}{r(\uptau-r-1)}+\ln(r+2)+\frac{5r+6}{2r(r+2)}\Big) \quad \mbox{if $\uptau<r-1$}\, , \cr
\frac{4\pi K}{3}\left(\ln (-\uptau)+\ln(r+2)-\ln 2-\frac{r}{2(r+2)}\right)\qquad \quad \mbox{if $\uptau>r-1$} \, .		
\end{cases}
																	\label{3.1.22}
\end{equation}
Expressing this in terms of $t$ using $\uptau=-e^{-t}$ and $T\equiv 3/(4\pi K)$
we get
\begin{equation}
P_1(t)=\begin{cases}
\displaystyle (e^{-t}+r+1)^{-\frac{1}{T}}(r+2)^{\frac{1}{T}}
\exp\left[\displaystyle 
-\frac{t}{T}+\frac{1}{T}\left\{-\frac{e^{-t}}{2r}-\frac{2(r+1)}{r(e^{-t}+r+1)}+\frac{5r+6}{2r(r+2)}\right\}\right] \cr
\hskip 2.8in \quad \mbox{for $t<-\ln(1-r)$}\, , \cr\cr
\displaystyle 
\left(\frac{r+2}{2}\right)^{\frac{1}{T}}\exp\left[\displaystyle
-\frac{t}{T}-\frac{r}{2 T(r+2)}\right]\qquad \quad \mbox{for $t>-\ln(1-r)$} \, .		
\end{cases}														\label{3.1.24}
\end{equation}
The same expression holds  for the survival probability $P_2(t)$ of $C_2$.
From this we
can find the life expectancy of $C_1$
\be \label{edefbart1}
\bar t_1 = \int_0^\infty P_1(t) dt\, .
\ee
As in the 1+1 dimensional case, $\bar t_1$ is slightly larger than $T$ but this
is simply due to the choice of initial condition that both observers are assumed
to exist at $t=0$. In Fig.~\ref{fcomp} we have plotted the ratio $\bar t_1/T$
as a function of $T$ for various values of $r$, and as we can see the result
remains close to 1. \changes{More discussion on $\bar t_1$ can be found below
\refb{ephi1series}.}

Next we consider the generalization of \refb{efirst}-\refb{ethird}. The
analysis is straightforward and we get the results
\ben \label{e4.13}
\frac{\partial\ln P_{12}(\uptau_1,\uptau_2)}{\partial\uptau_1}&=& 0 \qquad \mbox{for} \quad \uptau_1<\uptau_2-r\, ,  \nonumber \\
\frac{\partial\ln P_{12}(\uptau_1,\uptau_2)}{\partial\uptau_1}&=&\frac{2\pi K}{3}\left[\frac{2}{\uptau_1}+\changes{
{(r - \uptau_1+\uptau_2)(3 r - \uptau_1 - 3\uptau_2)\over r\, (r-\uptau_1-\uptau_2)^2}}
\right] \qquad \mbox{for} \quad \uptau_2-r<\uptau_1<\uptau_2+r
\nonumber \\ 
\frac{\partial\ln P_{12}(\uptau_1,\uptau_2)}{\partial\uptau_1}
&=& \frac{4\pi K}{3\uptau_1} \qquad \mbox{for} \quad \uptau_2+r<\uptau_1<0\, . 
\een
The solution to these equations, subject to the boundary condition
$\changep{P_{12}(\uptau_1=-1,\uptau_2)}=P_2(\uptau_2)=P_1(\uptau_2)$ is given by
\begin{equation}
\ln P_{12}(\uptau_1,\uptau_2)=\begin{cases}
\frac{4\pi K}{3}\left[\ln(-\uptau_2)+\ln(r+2)-\frac{r}{2(r+2)}-\ln 2\right]\quad  \mbox{if $\uptau_1<\uptau_2-r$} \\
\changed{\frac{4\pi K}{3}\Big[\ln (-\uptau_1)+\ln (-\uptau_2)-\ln (-\uptau_1-\uptau_2+r)+\frac{\uptau_1+\uptau_2}{2r}-\frac{2\uptau_1\uptau_2}{r(\uptau_1+\uptau_2-r)}}
\\ \changed{\hskip .75in  +\ln(r+2)+\frac{1}{r+2}\Big] 
\hskip 1in
\mbox{if $ \uptau_2-r<\uptau_1<\uptau_2+r$}} \\
\frac{4\pi K}{3}\left[\ln (-\uptau_1)+\ln (r+2)-\frac{r}{2(r+2)}-\ln 2\right]\quad 
\mbox{if $ \uptau_2+r<\uptau_1<0$}
\end{cases} 
\end{equation}
In terms of the physical time,  and $T=3/ (4\pi K)$, this becomes
\begin{equation}
P_{12}(t_1,t_2)=\begin{cases}
\{(r+2)/2\}^{1/T}
\exp\left[\displaystyle -\frac{r}{2T(r+2)}-\frac{t_2}{T}
\right]
\quad & \mbox{if \, $t_1<-\ln(r+e^{-t_2})$}\, \cr\cr
(r+2)^{1/T} (e^{-t_1}+e^{-t_2}+r)^{-\frac{1}{T}} \cr \quad \times
\exp\bigg[\displaystyle\frac{1}{T(r+2)}-\frac{1}{T}(t_1+t_2)-\frac{1}{2Tr}
& \hskip -.25in  \displaystyle  (e^{-t_1}
+e^{-t_2})
+\frac{2}{Tr}\frac{1}{e^{t_1}+e^{t_2}+re^{t_1+t_2}}
\bigg]
\\
%\qquad \qquad \qquad \qquad \qquad \qquad \qquad %\qquad  
& \hskip -.3in \mbox{if \, $-\ln(r+e^{-t_2})<t_1<-\ln(e^{-t_2}-r)$} \, \cr\cr
\{(r+2)/2\}^{1/T}\exp\left[\displaystyle -\frac{r}{2T(r+2)}-\frac{t_1}{T}\right]
\quad & \mbox{if \, $t_1>-\ln(e^{-t_2}-r)$}
\end{cases}
\label{prob}																	
\end{equation}
This gives 
\be \label{epp12}
\displaystyle P_{12}(t,t)
= (r+2)^{1/T}e^{\frac{1}{T\, (r+2)}} (2e^{-t}+r)^{-\frac{1}{T}}
\exp\left[-\frac{2}{T}t-\frac{1}{T\, r}e^{-t}+
\frac{2}{T\, r}\frac{1}{2e^{t}+re^{2t}}\right] \, .
\ee
In terms of this, and the functions $P_1=P_2$ given in
\refb{3.1.24}, we can calculate the
\changed{probability $\wt P_{12}$ 
of at least one of the two objects surviving} till time $t$ using
\be \label{ep12four}
\wt P_{12}(t) = P_1(t) + P_2(t) - P_{12}(t,t) 
\ee
and the combined life expectancy of two
objects using the analog of \refb{eavlife}
\be \label{et12four}
\bar t_{12} =\int_0^\infty \wt P_{12}(t) dt
= \int_0^\infty dt \left\{ P_1(t) + P_2(t) - P_{12}(t,t)\right\}
= 2\, \bar t_1 -  \int_0^\infty dt \, P_{12}(t,t)\, .
\ee

For the integral of $P_{12}(t,t)$
one can write down an expression in terms of special functions as follows.
Defining $y$ via
\begin{eqnarray}
2+re^{t}=\frac{2+r}{y}
\end{eqnarray}
for $r\ne 0$, we get
\be
\changed{\int_{0}^\infty dt\;P_{12}(t,t)
=[r(r+2)^{-1}]^{1/T}e^{\frac{1}{(r+2)T}}\int_0^{1} dy\; 
y^{-1+2/T}\left(1-\frac{2y}{2+r}\right)^{-1-1/T} \exp\left[{-\frac{y}{(2+r)T}}\right]\, .
\label{conflu1}}
\ee
Now, using the result
\begin{eqnarray} \label{edefphi1}
\int_0^1dy\; \frac{y^{a-1}(1-y)^{c-a-1}}{(1-u \, y)^b}e^{v \, y}=
B(a,c-a)\Phi_1(a,b,c;u,v)
\end{eqnarray}
with $Re\, c>Re\, a>0, |u|<1$,
$B$ the beta function and $\Phi_1$ the confluent hypergeometric series of two variables \changed{(Humbert series)}, we get
\be \label{ep12fin}
\int_{0}^\infty dt\;P_{12}(t,t)=
\frac{T}{2}\left[\frac{r}{(r+2)}\right]^{1/T}e^{\frac{1}{T(r+2)}}\;\;\Phi_1\left(
\frac{2}{T},1+\frac{1}{T},1+\frac{2}{T};\frac{2}{2+r},-\frac{1}{(2+r)T}\right)\, .
\ee
$\Phi_1$ has a power series expansion
\begin{eqnarray} \label{ephi1series}
\Phi_1(a,b,c;u,v)=\sum\limits_{m,n=0}^\infty \frac{(a)_{m+n}(b)_{m}}
{(c)_{m+n}m!n!}u^m v^n\, ,\quad |u|<1
\end{eqnarray}
where $(a)_m\equiv a(a+1)\cdots (a+m-1)$.

Unfortunately we have not been able to find an expression
for $\bar t_1=\int_0^\infty dt\, P_1(t)$ in terms of special functions. 
However, we can write down a series
expansion for this that will be suitable for studying its behaviour for small $r$.
The integral
of \refb{3.1.24} from $t=-\ln(1-r)$ to $\infty$ is
straightforward and yields
\be \label{esect1b}
T\, (1-r)^{1/T} (r+2)^{1/T} 2^{-1/T} \exp\left[-{r\over 2 T (r+2)}\right]\, .
\ee
The integral of  \refb{3.1.24} from $t=0$ to $-\ln(1-r)$ can be analyzed
by making a change of variable from $t$ to $y$ via
$e^{-t} = (1-y\, r)$. In terms of this variable the integral can be expressed
as
\be \label{efirt1b}
r\, \int_0^1 \, dy \, \left(1 - {yr\over r+2}\right)^{-1/T} (1-yr)^{-1+1/T}
\exp\left[-{2y^2 r (r+1)\over T(2+r)^3}\left(1-{yr\over 2+r}\right)^{-1}\right]
\exp\left[{y r^2\over 2T(2+r)^2}
\right]\, .
\ee 
Using series expansion of
the second and third terms in the integrand we get
\ben
&& \sum_{m,n=0}^\infty {1\over m! n!} 
\left(1-{1\over T}\right)_m (-1)^n {2^n r^{m+n+1} (r+1)^n\over T^n (2+r)^{3n}}
\nonumber \\
&& \hskip 1in \int_0^1 dy\, y^{m+2n} \left(1- {y r\over 2+r}\right)^{-n-1/T}
\exp\left[{y r^2\over 2 T (2+r)^2}\right]\, .
\een
The integral over $y$ can be expressed in terms of 
$\Phi_1$ using
\refb{edefphi1}. Adding \refb{esect1b} to this we get
\ben \label{et1fin}
\bar t_1 &=& T\, (1-r)^{1/T} (r+2)^{1/T} 2^{-1/T} \exp\left[-{r\over 2 T (r+2)}\right]
\nonumber \\
&& +  \sum_{m,n=0}^\infty {1\over m! n!} {1\over m+2n+1} 
\left(1-{1\over T}\right)_m (-1)^n {2^n r^{m+n+1} (r+1)^n\over T^n (2+r)^{3n}}
\nonumber \\
&& \hskip 1in
\Phi_1\left(m+2n+1, n + {1\over T}, m+2n+2; {r\over 2+r}, {r^2\over 2T (2+r)^2}
\right)\, .
\een

It can be checked using \refb{3.1.24}, \refb{edefbart1}, 
\refb{epp12} and \refb{et12four} that for $r\to 0$ we get 
$\bar t_{12}/\bar t_1=1$ and for $T\to\infty$ we get 
$\bar t_{12}/\bar t_1=3/2$.
The values of $\hbox{\rm `gain'} \equiv \bar t_{12}/\bar t_1
$ for different values of $r$ have been
plotted against $T$ in Fig.~\ref{fnum}.

\subsection{The case of small initial separation}

\begin{figure}
\begin{center}
\epsfbox{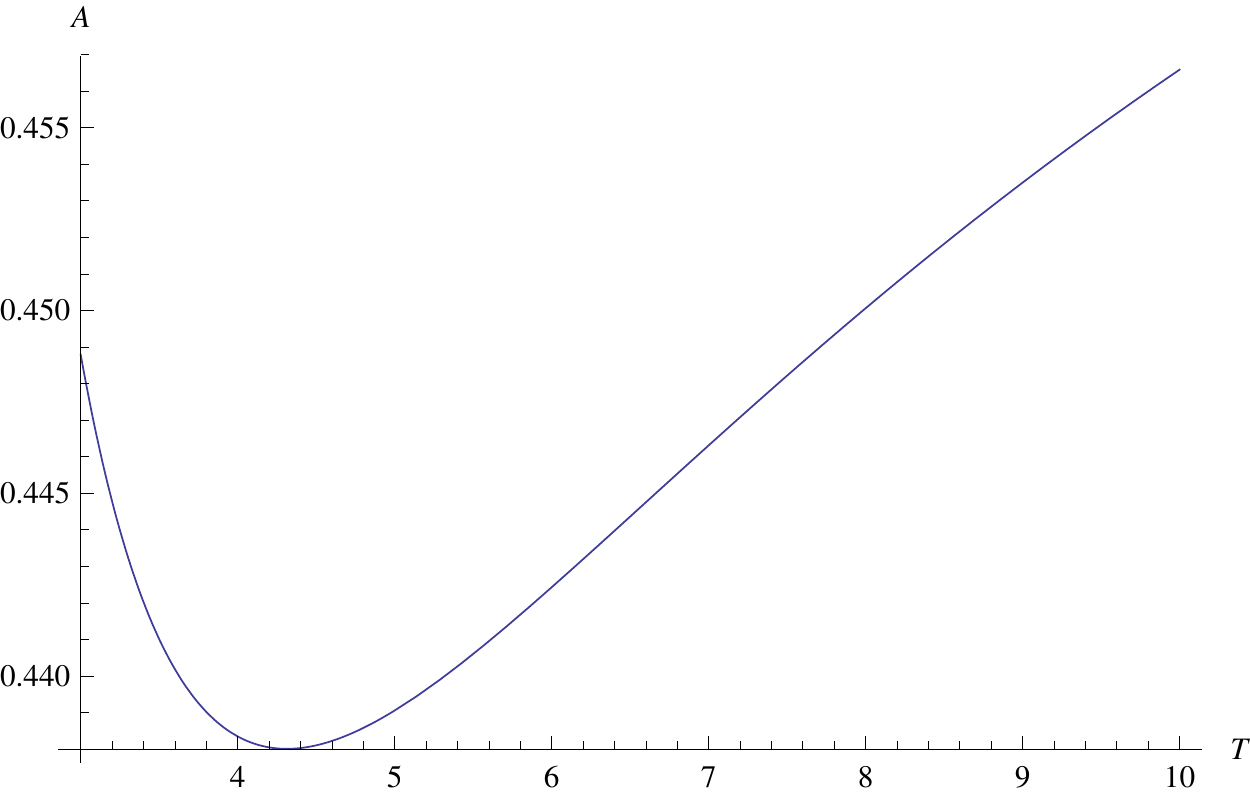}
\end{center}
\caption{\changec{The coefficient of the $r^{1/T}$ term in the 
expression for $\bar t_{12}
/\bar t_1$ as a function of $T$}. \label{fat}}
\end{figure}

Since from practical considerations the small $r$ region is of interest,
it is also useful to consider the expansion of $\bar t_{12}/\bar t_1$ 
for small $r$. For this we have to analyze the behaviour of $\bar t_1$ as
well as that of $\int_0^\infty dt\, P_{12}(t,t)$ for small $r$. Let us
begin with $\bar t_1$ given in \refb{et1fin}. 
It can be easily seen that this is given by \changec{$T+\OO(r)$} with the 
contribution $T$ coming from the first term. However, the contribution
from $\int_0^\infty dt\, P_{12}(t,t)$ has a more complicated behaviour
at small $r$. This is related to the fact that in the $r\to 0$ limit the fourth
argument of $\Phi_1$ in \refb{ep12fin} approaches 1, and in this limit
the series expansion \refb{ephi1series} diverges. To study the
small $r$ behaviour we shall go back to the original expression
for $P_{12}(t,t)$ given in \refb{epp12}.
We change variable to $v=e^{-t}/r$ and write
\ben
\int_0^\infty dt\, P_{12}(t,t) &=& (r+2)^{1/T} e^{1 / T(r+2)} r^{1/T} 
\int_0^{1/r} {dv\over v} (2v+1)^{-1/T} v^{2/T} \exp\left[-{v\over T(2v+1)}\right]
\nonumber \\
&=& (r+2)^{1/T} e^{1 / T(r+2)} r^{1/T} 
\int_0^{1/r} {dv\over v}  v^{2/T} \left[
(2v+1)^{-1/T} \exp\left[-{v\over T(2v+1)}\right]\right.
\nonumber \\  && \left. - (2v)^{-1/T}
\exp\left[-{1\over 2T}\right]\right] + (r+2)^{1/T} 2^{-1/T}  
\exp\left[{1 \over T(r+2)}
-{1\over 2T}\right] T\, , \nonumber \\
\een
where in the \changek{last} step we have subtracted an integral from the original
integral and compensated for it by adding the explicit result for the integral.
This subtraction makes the integral convergent even when we replace the
upper limit $1/r$ by $\infty$.
Taking the small $r$ limit we get
\ben
\int_0^\infty dt\, P_{12}(t,t) 
&=&  2^{1/T} r^{1/T}\int_0^\infty {dv\over v}  v^{2/T} \left[
(2v+1)^{-1/T} \exp\left[-{v\over T(2v+1)} +{1\over 2T}\right]
- (2v)^{-1/T}
\right] \nonumber \\ &&
+ \, T + \OO(r)\, .
\een
Combining this with the earlier result that
for $r\to 0$, $\bar t_1\simeq T + \changec{\OO(r)}$ 
and using \refb{et12four} we get
\be
{\bar t_{12}\over \bar t_1} = 2 - {1\over \bar t_1} \int_0^\infty dt\, P_{12}(t,t)
= 1 + A(T)\, r^{1/T}\, ,
\ee
where 
\be \label{edefA}
A(T) = T^{-1} 2^{1/T} \int_0^\infty dv \, v^{-1+2/T} \left[(2v)^{-1/T} -
(2v+1)^{-1/T} \exp\left(-{v\over T (2v+1)}+{1\over 2T}\right)
\right]\, .
\ee
The numerical values of $A(T)$ are moderate
-- for example \changed{$A(5)\simeq 0.439$ and $A(10)\simeq 0.457$}.
\changec{A plot of $A(T)$ as a function of $T$ has been shown in 
Fig.~\ref{fat}.}
The $1/T$ exponent of $r$ 
shows that even if we begin with small $r$, for 
moderately large $T$  (say $T\sim 5$) we can get moderate
enhancement in life expectancy.

\sectiono{Generalizations} \label{sgen}

In this section we shall discuss various possible generalizations of our
results. 

\subsection{Multiple objects in de Sitter space} \label{smultiple}

We shall begin by discussing the case of three objects 
$C_1$, $C_2$ and $C_3$ placed at certain
points in $3+1$ dimensional de Sitter space-time and analyze the probability
that at least one of them will survive till time $t$. Let $P_{123}(t_1,t_2,t_3)$
denote the probability that \changes{$C_1$ survives till time $t_1$, 
$C_2$ survives till time $t_2$ and $C_3$} survives till time $t_3$. 
Similarly $P_{ij}(t_i,t_j)$ for $1\le i,j\le 3$ will denote the probability that $C_i$
survives till $t_i$ and $C_j$ survives till
$t_j$ and $P_i(t_i)$ will denote the probability that $C_i$ survives
till $t_i$. All probabilities are defined under the prior assumption that all
\changen{objects} are alive at $t=0$. These probabilities can be calculated by
generalizing the procedure described in \S\ref{s11} and \S\ref{s31} by 
constructing ordinary differential equations in one of the arguments at fixed
values of the other arguments.  The geometry of course now becomes 
more involved due to the fact that the past light cone of one object will
typically intersect the past light cones of the other objects which
themselves may have overlaps, and
one has to carefully subtract the correct volume. But the analysis is
straightforward.

\begin{figure}
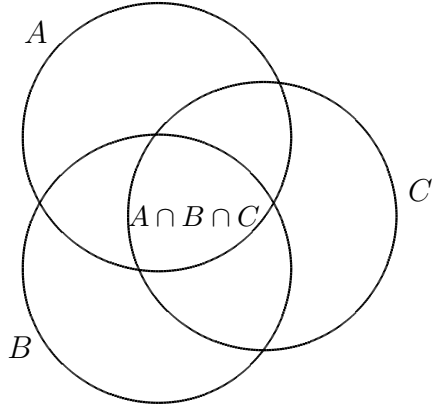


%\vskip -.2in

\begin{center}
\figvennthree
\end{center}

\vskip -.4in

\caption{The Venn diagram \changen{illustrating} 
that the survival probability of one
of $A$, $B$ or $C$, denoted by $P(A\cup B\cup C)$, is given by
$P(A)+P(B)+P(C)-P(A\cap B)-P(B\cap C)-P(A\cap C)+P(A\cap B\cap C)$.
} \label{figvenn3}

\end{figure}

The quantity of direct \changes{interest  is} the probability $\wt P_{123}(t)$
that at least one of the objects survives till time $t$. With the help of the
Venn diagram given in Fig.~\ref{figvenn3} we get
\be
\wt P_{123}(t) = \Big( P_{1}(t)+P_{2}(t)+P_{3}(t)
-P_{12}(t,t)- P_{13}(t,t)-P_{23}(t,t)+P_{123}(t,t,t) \Big)
\, .
\ee
Using this we can calculate the life expectancy of the combined system as
\be
-\int_0^\infty \, dt \, t \, \changen{{d\over dt} \wt P_{123}(t)} 
= \int_0^\infty dt \, \wt P_{123}(t)\, .
\ee

The generalization to the case of $N$ objects is now obvious. The
relevant formula is
\be \label{efin1}
\wt P_{12\cdots N}(t) = 
\Big(\sum_{i=1}^{N}P_i(t)-\sum_{i<j}^{N}P_{ij}(t,t)+ \sum_{i<j<k}^{N}P_{ijk}(t,t,t)
+ \cdots (-1)^{N+1}P_{12\cdots N}(t,t,\cdots,t)\Big)
\ee
where $P_{i_1\cdots i_k}(t_{i_1},\cdots t_{i_k})$ are again computed by
solving ordinary differential equations in one of the variables. 
Once $\wt P_{12\cdots N}(t)$ is computed we can get the life expectancy
of the combined system by using
\be \label{efin2}
\bar t_{12\cdots N} = \int_0^\infty \wt P_{12\cdots N}(t)  dt\, .
\ee

\begin{figure}
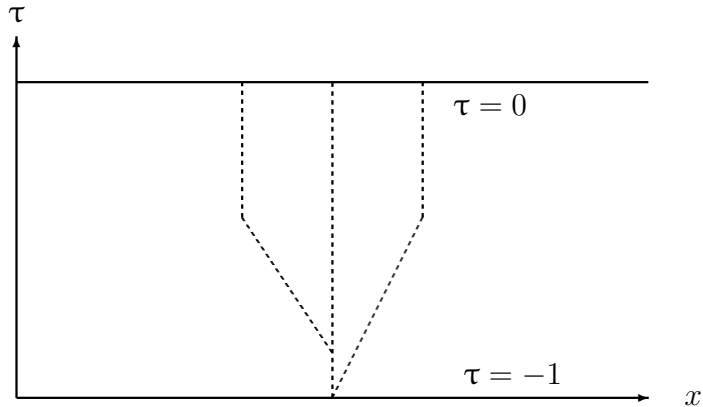

\begin{center}
\figmulti
\end{center}
\caption{Multiple objects originating from the same space-time point.
Different dashed lines represent the trajectories followed by different 
objects.} 
\label{figmulti}
\end{figure}

\subsection{Realistic trajectories} 
\label{sjourney}

Another generalization involves considering a situation where multiple objects
originate at the same space time point and then follow different trajectories,
eventually settling down at different comoving coordinates. This has been 
illustrated in Fig.~\ref{figmulti}. \changes{This represents} 
the realistic situation
since by definition different civilizations of the same race must originate at 
some common source. 
We can now generalize our analysis to take into account the possibility of
decay during the journey as well.
Eqs.\refb{efin1} and \refb{efin2} still holds, but the
computation of $P_{i_1\cdots i_k}(t_{i_1},\cdots t_{i_k})$ will now have to
be done by taking into account the details of the trajectories of each object
and the overlaps of their past light cones. The principle remains the same, and
we can set up ordinary differential equations for each of these quantities. The
only difference is that the \changes{spatial}
separation between the $i$-th object at
$\uptau=\uptau_i$ and the $j$-th object at $\uptau=\uptau_j$ will now depend
\changek{on} $\uptau_i$ and $\uptau_j$ according to the trajectories followed by them.

This analysis can be easily generalized to the case where 
each of the descendant  objects
in turn produces its own descendants \changek{which  settle away from the parent
object and
eventually 
go outside each other's horizon due to the Hubble expansion}. 
If this could be repeated at a rate 
faster than the
vacuum decay rate then we can formally ensure that some of the objects
will survive vacuum decay\cite{1503.08130}. 
However, since within a few Hubble periods
most of the universe will split up into gravitationally bound systems outside
each other's horizon, in practice 
this is going to be an increasingly difficult task.

\subsection{Matter effect} \label{smatter}

A third generalization will involve relaxing the assumption that 
the universe
has been de Sitter throughout its past history. While
de Sitter metric will be a good 
approximation after a few Hubble periods, within
the next few Hubble periods 
we shall still be sensitive to the fact that the universe had
been matter dominated in the recent past and had a beginning. This will
change the form of the metric \refb{eflat} to
\be 
ds^2 = -dt^2 + a(t)^2 (dx^2+dy^2+dz^2)
\ee
where $a(t)$ is determined from
the Friedman equation 
\be \label{edota}
{1\over a} {da\over dt}= \sqrt{{8\pi G\over 3} \left(\rho_\Lambda + {\rho_m 
\over a^{3}}\right)}
\ee
in the convention that the value of $a$ is 1 today and $\rho_\Lambda$
and $\rho_m$ are the energy densities due to cosmological constant and
matter today. 
Since we have chosen the  unit of time so that the Hubble parameter in the
cosmological constant dominated universe is 1, we have
$\sqrt{8\pi G\rho_\Lambda/3}=1$. Defining\footnote{\changek{We use 
cosmological
parameters given in \cite{1502.01589}}.}
\be
\changed{c\equiv \rho_m/\rho_\Lambda \simeq 0.45}
\ee
we can express \refb{edota} as
\be
{1\over a} {da\over dt}= \sqrt{1 + c a^{-3}}\, .
\ee
Let $\uptau$ be the conformal time defined via
\be
d\uptau = dt / a(t)
\ee
with the boundary condition $\uptau\to 0$ as $t\to\infty$. 
Then \refb{edota} takes the form
\be
{1\over a^2} {da\over d\uptau} = \sqrt{1 + c a^{-3}}\, ,
\ee
whose solution is
\be \label{etaum}
\uptau = \changek{-\int_a^\infty {d  b\over b^2 \sqrt{1 + c b^{-3}}} }
=-\int_0^{1/a} {d v\over \sqrt{1 + cv^3}}
= - {1\over a} ~_2F_1\left({1\over 3}, {1\over 2}; {4\over 3}, 
-{c\over a^3}
\right)\, .
\ee
This implicitly determines $a$ as a function of $\uptau$. The metric is
given by
\be 
ds^2 = a(\uptau)^2 (-d\uptau^2 + dx^2 + dy^2 + dz^2)\, .
\ee

Using the experimental value \changed{$c\simeq 0.45$} we get that 
$\uptau\to \uptau_0\simeq -3.7$ as $a\to 0$, 
showing that the big bang singularity is at $\uptau\simeq
-3.7$.\footnote{Of course close to the singularity the universe becomes
radiation dominated but given the short span of radiation dominated era
we ignore that effect for the current analysis.}  We also have that at
$a=1$, $\uptau\simeq -0.95$.
This is not very different from the value $\uptau=-1$ for pure de Sitter 
\changes{space-time} with
which we have worked. 
\changek{However, we shall now show that}
the decay rate in the matter dominated epoch of the universe
differs significantly from that in the cosmological constant dominated
epoch. \changek{The decay rate} of an isolated observer at some
value of the conformal time $\uptau$ is
given by the following generalization of \refb{e4.2}, \refb{e4.3}:
\ben \label{etoday0}
{d\over d\uptau}\ln P_0(\uptau)
&=& \changec{- 4\pi K\int_{\uptau_0}^{\uptau}  d\sig\, a(\sig)^4 \, 
(\uptau - \sig)^2 }
\nonumber \\
&=& \changec{-4\pi K \int_{b=0}^{a(\uptau)} {db\over b^2\sqrt{1+ cb^{-3}}} \, b^4 \, 
\, \left\{ \uptau + {1\over b} ~_2F_1\left({1\over 3}, {1\over 2}; {4\over 3}, 
-{c\over b^3}
\right)\right\}^2} \nonumber \\
\een
where in the second step we have changed the integration variable from
$\sig$ to $b=a(\sig)$.  From this we can compute the decay rate:
\ben \label{egeneral}
D(t) &\equiv & 
-{d\over d t}\ln P_0=  -{1\over a(t)} 
{d\over d\uptau}\ln P_0(\uptau) \nonumber \\
&=& {4\pi K \over a(t)}
\int_{b=0}^{a(t)} {db\over \sqrt{1+ cb^{-3}}} \, b^2 \, 
\, \left\{ \uptau(t) + {1\over b} ~_2F_1\left({1\over 3}, {1\over 2}; {4\over 3}, 
-{c\over b^3}
\right)\right\}^2\, .
\een
 Using the information that today $a=1$ and
$\uptau\simeq -0.95$ we get
\be \label{etoday}
D(t)|_{\rm today} \simeq {4\pi K\over 3} \times 0.067\simeq 
 {0.067\over T}\, .
 \ee
This is lower than the corresponding rate $T^{-1}$ in the de Sitter
epoch by about a factor of \changed{15.}  The growth 
of the decay rate with scale factor has been shown in Fig.~\ref{frate}.

Our analysis of \S\ref{s31} can now be
repeated for two or more observers and also for general
trajectory discussed in \S\ref{sjourney} 
with this general form of the
metric to get more accurate computation of the life expectancy. These
corrections will be important if $T\lsim 1$ and the decay takes place 
within a few Hubble period from now. On the other hand if $T$ is large
(say $\gsim 10$) then the decay is likely to take place sufficiently far 
in the future by which time the effect of our matter dominated past
will have insignificant effect on the results.

The fact that the decay rate increases with time
 till it eventually
settles down to a constant value in the de Sitter epoch has some important
consequences:
\begin{enumerate}
\item
We have already seen from \refb{etoday} that the decay
rate today is about \changed{15} times smaller than the decay rate in the
de Sitter epoch.  \changek{Eq.\refb{egeneral} for $a(t)=2$} shows that
even when the universe will be double its size compared to today, the decay
rate will remain at about 
\changed{27\%} of the decay rate in the de Sitter epoch.
Since most of the journeys to different parts of the universe -- if they take place
at all -- are likely to happen 
during this epoch, we see that the probability of decay
during the journey will be considerably less than that in the final de Sitter
phase. This partially justifies our analysis in \S\ref{s31} where we neglected
the probability of decay during the journey. This also shows that if we 
eventually carry out a detailed numerical analysis taking into account
the effect discussed in \S\ref{sjourney}, it should be done in conjunction 
with the analysis of this subsection taking into account the effect of 
matter.

\item 
It is also possible to see \changek{from \refb{egeneral}, \refb{etoday}
(or Fig.~\ref{frate})} 
that the decay
rate in the past was even smaller than that of \changek{today. 
If $D(t)$ denotes} the decay rate at time $t$ defined 
in \refb{egeneral},
then the average decay rate in our past 
can be defined as
\be \label{ez1}
{1\over t_1} \int_0^{t_1} D(t) dt\, ,
\ee
where $t_1$ denotes the current age of the universe given by
\be \label{ez2}
t_1=\int_0^1 da\, \left({da\over dt}\right)^{-1} =  
\int_0^1 {da \over a\sqrt{1+ca^{-3}}} \changed{\simeq 0.79 \, .}
\ee
In physical units $t_1$ is about \changek{$1.38\times 10^{10}$ years.
The evaluation} of \refb{ez1} can be
facilitated using the observation that 
$D(t) dt$ is $K$ times the volume enclosed between the past light cones
of the object at times $t$ and $t+dt$. Therefore, $\int_0^{t_1} dt D(t)$
must be $K$ times 
the total volume enclosed by the past light cone of the 
object at $t_1$. This can be easily computed, yielding
\ben \label{etoday1}
\int_0^{t_1} D(t) dt &=& {4\over 3}\, \pi\, K 
\int_{\uptau_0}^{\uptau}  d\sig\, a(\sig)^4 \, (\uptau - \sig)^3 
\nonumber \\
&=& {1\over T} \, \int_{b=0}^{a(\uptau)} {db\over b^2\sqrt{1+ cb^{-3}}} \, b^4 \, 
\, \left\{ \uptau + {1\over b} ~_2F_1\left({1\over 3}, {1\over 2}; {4\over 3}, 
-{c\over b^3}
\right)\right\}^3 \, .
\een
For \changed{}$c\simeq 0.45$ 
and $\uptau$ given by today's value $-0.95$ this gives
\be  \label{etoday2}
\int_0^{t_1} D(t) dt  = \changed{{0.014\over T}}\, .
\ee
Using \refb{ez1}, \refb{ez2} we get the average decay rate to be 
\changed{$0.018/T$}. This is about \changes{$3.7$} 
times smaller than the present decay rate 
given in \refb{etoday} and \changed{56} 
times smaller than the decay rate $1/T$
in the de Sitter epoch.

Integrating the equation \changep{$dP_0/dt = - D(t) P_0(t)$} 
we get
\be
\ln P_0 (t_1) = - \int_0^{t_1} dt D(t) = -{0.014\over T}\, .
\ee
Requiring this to be not much smaller than $-1$ (which is equivalent
to requiring that the inverse of the average decay rate
\refb{ez1} be not much smaller than 
the age of the universe $t_1$) gives
$T\gsim 0.014$. This is much lower than what one might have naively
predicted by equating the lower bound \changek{on} $T$ 
to the age of the universe i.e. 
$T\gsim t_1\sim 0.79$. 
Recalling that the unit of time is set by the Hubble period in the de Sitter
epoch which is about $1.7\times 10^{10}$ years, the bound $T\gsim 0.014$
translates to a lower bound
of order $2.5 \times 10^8$ years. Since the current decay rate is about 
15 times smaller than that in the
final de Sitter epoch, we see that the lower bound on the current inverse
decay rate is of order $3.7\times 10^9$ years. This is comparable to the
period over which \changek{the}
earth is expected to be destroyed due to the expanded
size of the Sun.

\item Finally we note that the above analysis was based on the assumption that
the bubbles continue to nucleate and expand in the FRW metric
at the same rate as
they would do
in the \changek{metastable} 
vacuum. This will be expected as long as the \changek{matter and radiation
density and temperature are}
small compared to the microscopic scales involved in the
bubble nucleation process, {\it e.g.} the scale set by the negative cosmological
constant of the vacuum in the interior of the bubble. Some discussion on the
effect of cosmological space-time background on the bubble 
nucleation / evolution can be found in \cite{0711.3417,0908.2757}.  

\end{enumerate}

\sectiono{Discussion} \label{sdiss}

We have seen that the result for how much we can increase the life expectancy
by spreading out in space depends on the parameters $r$ and $K$,
which in turn are 
determined by the Hubble parameter of the de Sitter
space-time, the initial spread between different objects and the 
\changem{inverse decay rate}. Therefore, the knowledge of these
quantities is important for planning our future course of action if we are to
adapt this strategy for increasing the life expectancy of the human race. In this
section we shall discuss possible strategies for determining /
manipulating these quantities.

We begin with the Hubble expansion parameter  $H$. This is determined by 
the cosmological constant which has been quite well measured by now.
Assuming that the current expansion rate 
is of order 68Km/sec/Mpc and accounting for the fact that the cosmological
constant accounts for about \changed{69\%} of the total energy density
we get $H^{-1}\simeq 1.7\times 10^{10}$ years.
Future experiments will undoubtedly provide a more accurate determination
of this number, but given the uncertainty in the other quantities, this will
not significantly affect our future course of action. Of course we may discover
that the dark energy responsible for the accelerated expansion of the 
universe comes from another source, in which case we have to reexamine the
whole situation.

Next we turn to the initial separation between different objects which
determine the value of $r$. Since in order for the \changek{Hubble} 
expansion to be
effective in separating the objects they have to be unbound 
gravitationally, a minimum separation between  the objects
is necessary for overcoming
the attractive gravitational force of the home galaxy. 
For example the size of our local gravitationally bound group of galaxies
is of order 5 million light years which correspond 
to $r\sim 3 \times 10^{-4}$. The question is whether larger values
of $r$ can be accessed. An interesting analysis by Heyl\cite{0509268} 
concluded
that by building a space-ship that can constantly accelerate /
decelerate at a value
equal to the acceleration due to gravity, we can reach values of $r$ close 
to unity in less than 100 years viewed from the point of view of the
space-traveller. Of course this will be close to about $10^{10}$ years 
viewed from earth, and roughly the reduction of time viewed from the 
space-ship can be attributed to the large time dilation \changed{at} 
the peak
speed of the space-ship reaching a value close to that of light. However, this
large time dilation will also increase the effective temperature of the microwave
background radiation in the forward direction and without a proper shield such
a journey will be impossible to perform. If one allows a maximum
time dilation of
the order of 100 then the microwave temperature in the forward direction 
rises to about the room temperature. Even then we have to worry about
the result of possible collisions with intergalactic dust and othe debris in space.
Even if these problems are resolved, 
we shall need a time of order $10^8$ years
from the point of view of the space-ship to travel a distance of order
$10^{10}$ light years. Even travelling the minimum required distance 
of order $10^7$ light years will take $10^5$ years in such a space-ship.
Such a long journey in a space-ship 
does not seem very practical but may not be impossible.

Another interesting suggestion for populating regions of space-time
which will eventually be outside each other's horizon has been made
by Loeb\cite{1102.0007}. Occasionally there are hypervelocity stars which escape 
our galaxy (and the cluster of galaxies which are gravitationally bound)
and so if we could find a habitable planet in such a star we could take
a free ride in that planet and escape our local gravitationally bound
system. In general of course
there is no guarantee that such a star will reach another cluster of galaxies
where we could spread out and thrive, but some time we may be lucky. 
It has been further suggested in \cite{1411.5022,1411.5030} 
that the merger of Andromeda
and the Milky Way galaxies in the future\cite{0204249} \changed{could}
generate a large number of
such hypervelocity 
stars travelling at speeds comparable to \changed{that} of light and they
could travel up to \changek{distances of the order of}
$10^9$ light years by the time they burn out.
This could allow us to achieve values of $r$ of order $10^{-1}$ or more.

Let us now turn to the value of $T$ or equivalently the decay rate of 
the de Sitter vacuum in which we currently live. 
This is probably the most important ingredient since we have seen that
for $T\lsim 1$ we do not gain much by spreading out, while for large
enough $T$ we can achieve the maximum possible gain, given by the
harmonic numbers, by spreading out even over modest distances of
$r\sim 10^{-3}$. At the same time if $T$ is so large that it exceeds the
\changed{period  over} which galaxies will die
then vacuum decay \changed{may} 
not have a significant role in deciding our end 
and we should focus on other issues. For this reason
estimating the value of $T$ seems to be of paramount importance.
\changed{Unfortunately, due to the very nature of the vacuum decay process 
it is not possible to
determine it by any sort of direct experiment since such an experiment
will also destroy the observer. It may be possible in the future to device clever
indirect experiments to probe vacuum instability without actually causing
the transition to the stable vacuum, but no such scheme is known at
present.}

At a crude level 
the current age of the universe \changed{-- which is about $0.79$ times the
asymptotic Hubble period in the cosmological constant dominated epoch
--}
together with the assumption that
we have not been extremely lucky to survive this long, 
suggests that the \changem{inverse decay rate} of the 
universe is \changed{$\gsim 0.8$}. However, \refb{etoday1} shows that this
actually gives a lower bound of \changed{$T\gsim 0.014$} reflecting the fact that 
the decay rate in the de Sitter
epoch will be about \changed{56} times faster than the average decay rate in
the past. 
If we allow for the possibility that we might have been extremely lucky to
\changed{have survived} till today, then
we have indirect arguments that \changed{lower the bound by a factor of 
10\cite{0512204}.
Clearly these rates are} too fast and if $T$ really happens to be less than
1, then there is not much we can do to prolong our collective life.

Is there any hope
of computing $T$ theoretically? Unfortunately any bottom up approach based
on the analysis of low energy effective field theory is insufficient for this problem.
The reason for this is that the vacuum decay rate is a heavily ultraviolet sensitive
\changed{quantity. Given} 
a theory with a perfectly stable vacuum we can add to it a new
\changed{heavy} scalar field whose effect will be strongly suppressed at low
energy, but which can have a potential that makes the vacuum metastable with
arbitrarily large decay rate. For this reason the only way we could hope to
estimate $T$ is through the use of a top down approach in which we have a
fundamental microscopic theory all of whose parameters are fixed by some
fundamental principle, and then compute the vacuum decay rate using standard
techniques. In the context of string theory this will require finding the
vacuum in the landscape that describes our universe. Alternatively,
in the multiverse
scenario, we need to
carry out a statistical analysis that establishes that the overwhelming
majority of the vacua that resemble  our vacua will have their decay rate
lying within a narrow range. This can then be identified as the likely value
of the decay rate. There have been attempts in this 
direction\cite{0706.3201,0712.1397,0907.4153,1010.5241}, 
but it is probably
fair to say that we do not yet have a definite result based on which we can 
plan our future course of action.

{\bf Acknowledgement:}
\changem{We would like to thank  Adam Brown,
Rajesh Gopakumar and Abraham Loeb for
useful discussions}.
This work  was
supported in part by the 
DAE project 12-R\&D-HRI-5.02-0303. The work of A.S. was also 
supported in part by the J. C. Bose fellowship of 
the Department of Science and Technology, India and the KIAS 
distinguished professorship.
The work of \changed{MV} was also supported by the
SPM research grant of the Council for Scientific and Industrial Research (CSIR), India.

\appendix

\sectiono{Decay rate for equation of state $p=w\, \rho$} \label{sapp}

In this appendix we shall compute the growth of decay rate with time
for a general equation of state of the form $p=w\, \rho$ with $w>-1$. 
In this case
the $\rho$ and $a$ are related as
\be 
\rho = \rho_0 \, a^{-3(w+1)}\, ,
\ee
for some constant $\rho_0$. 
As a result the dependence of $a$ on $t$ and $\uptau$ are determined by the
equations
\be 
{1\over a} {da\over dt} = \sqrt{{8\pi G\over 3} \rho} 
= C\, a^{-3(w+1)/2}, \qquad C \equiv \sqrt{{8\pi G\over 3} \rho_0}\, ,
\ee
and
\be
{1\over a^2} {da\over d\uptau} = C\, a^{-3(w+1)/2}\, .
\ee
The solutions to these equations are
\be \label{esolt}
t = {2\over 3 C (w+1)} a^{3(w+1)/2}, \quad 
\uptau = {2\over (3 w+1) C} a^{(3w+1)/2}\, .
\ee
As $a$ ranges from 0 to $\infty$, both $t$ and $\uptau$ also range from 0 to
$\infty$.

We can now compute the decay rate using the first equation of \refb{etoday0}:
\ben \label{eappdt}
D(t) &\equiv& - {1\over a} {d\over d\uptau} \ln P_0(\uptau) 
\nonumber \\
&=& {1\over a} \, 4\pi K \, \int_0^\uptau d\sig\, a(\sig)^4 \, 
(\uptau - \sig)^2
\, ,
\een
where $a(\sigma)$ denotes the scale factor at conformal time $\sigma$.
Changing integration variable to $b=a(\sigma)$, which due to \refb{esolt}
corresponds to 
\be 
\sigma = {2\over (3 w+1) C} b^{(3w+1)/2}\, ,
\ee
we can express \refb{eappdt} as
\ben
D(t) &=& 4 \pi K \, a^{-1} \left({2\over (3 w+1) C}\right)^3 \, {3w+1\over 2}\, 
\int_0^a db \, b^4 \, b^{(3w-1)/2} \, \left( a^{(3w+1)/2)} - b^{(3w+1)/2}\right)^2
\nonumber \\
&=& {32 \pi K\over 3\, C^3} \, a^{9(w+1)/2} \, {1\over (w+3) (3w+5) (9w+11)}
\, .
\een
Using the relation between $a$ and $t$ given in \refb{esolt}, this can be
expressed as
\be \label{edtfull}
D(t) = 36\, \pi \, K \, {(w+1)^3\over (w+3) (3w+5) (9w+11)}\, t^3\, .
\ee


\begin{thebibliography}{99}

\bibitem{okun} 
  I.~Y.~Kobzarev, L.~B.~Okun and M.~B.~Voloshin,
  ``Bubbles in Metastable Vacuum,''
  Sov.\ J.\ Nucl.\ Phys.\  {\bf 20}, 644 (1975)
  [Yad.\ Fiz.\  {\bf 20}, 1229 (1974)].
  %%CITATION = SJNCA,20,644;%%

\bibitem{stone} 
  M.~Stone,
  ``The Lifetime and Decay of Excited Vacuum States of a Field Theory Associated with Nonabsolute Minima of Its Effective Potential,''
  Phys.\ Rev.\ D {\bf 14}, 3568 (1976).
  %%CITATION = PHRVA,D14,3568;%%
  %60 citations counted in INSPIRE as of 24 mar 2015

\bibitem{frampton} 
  P.~H.~Frampton,
  ``Consequences of Vacuum Instability in Quantum Field Theory,''
  Phys.\ Rev.\ D {\bf 15}, 2922 (1977).
  %%CITATION = PHRVA,D15,2922;%%
  %66 citations counted in INSPIRE as of 24 mar 2015

\bibitem{coleman1} 
  S.~R.~Coleman,
  ``The Fate of the False Vacuum. 1. Semiclassical Theory,''
  Phys.\ Rev.\ D {\bf 15}, 2929 (1977)
  [Erratum-ibid.\ D {\bf 16}, 1248 (1977)].
  %%CITATION = PHRVA,D15,2929;%%

\bibitem{coleman2} 
  C.~G.~Callan, Jr. and S.~R.~Coleman,
  ``The Fate of the False Vacuum. 2. First Quantum Corrections,''
  Phys.\ Rev.\ D {\bf 16}, 1762 (1977).
  %%CITATION = PHRVA,D16,1762;%%

\bibitem{coleman} 
  S.~R.~Coleman and F.~De Luccia,
  ``Gravitational Effects on and of Vacuum Decay,''
  Phys.\ Rev.\ D {\bf 21}, 3305 (1980).
  %%CITATION = PHRVA,D21,3305;%%

\bibitem{9805201} 
  A.~G.~Riess {\it et al.}  [Supernova Search Team Collaboration],
  ``Observational evidence from supernovae for an accelerating universe and a cosmological constant,''
  Astron.\ J.\  {\bf 116}, 1009 (1998)
  [astro-ph/9805201].
  %%CITATION = ASTRO-PH/9805201;%%
  
\bibitem{9812133} 
  S.~Perlmutter {\it et al.}  [Supernova Cosmology Project Collaboration],
  ``Measurements of Omega and Lambda from 42 high redshift supernovae,''
  Astrophys.\ J.\  {\bf 517}, 565 (1999)
  [astro-ph/9812133].
  %%CITATION = ASTRO-PH/9812133;%%

\bibitem{0004134} 
  R.~Bousso and J.~Polchinski,
  ``Quantization of four form fluxes and dynamical neutralization of the cosmological constant,''
  JHEP {\bf 0006}, 006 (2000)
  [hep-th/0004134].
  %%CITATION = HEP-TH/0004134;%%

\bibitem{0105097} 
  S.~B.~Giddings, S.~Kachru and J.~Polchinski,
  ``Hierarchies from fluxes in string compactifications,''
  Phys.\ Rev.\ D {\bf 66}, 106006 (2002)
  [hep-th/0105097].
  %%CITATION = HEP-TH/0105097;%%


\bibitem{0301240} 
  S.~Kachru, R.~Kallosh, A.~D.~Linde and S.~P.~Trivedi,
  ``De Sitter vacua in string theory,''
  Phys.\ Rev.\ D {\bf 68}, 046005 (2003)
  [hep-th/0301240].
  %%CITATION = HEP-TH/0301240;%%
  %2062 citations counted in INSPIRE as of 20 mar 2015
%\cite{Giddings:2001yu}

\bibitem{bostron}
N. Bostrom, 
``Anthropic Bias: Observation Selection Effects
in Science and Philosophy,'' Routledge: New York,
2002.

\bibitem{0512204} 
  M.~Tegmark and N.~Bostrom,
 ``How unlikely is a doomsday catastrophe?,''
  astro-ph/0512204.
  %%CITATION = ASTRO-PH/0512204;%%

\bibitem{1505.06397} 
A.~Masoumi, ``Topics in vacuum decay,''
 arXiv:1505.06397 [hep-th].
  %%CITATION = ARXIV:1505.06397;%%

\bibitem{guth} 
  A.~H.~Guth,
  ``The Inflationary Universe: A Possible Solution to the Horizon and Flatness Problems,''
  Phys.\ Rev.\ D {\bf 23}, 347 (1981).
  %%CITATION = PHRVA,D23,347;%%

\bibitem{1503.08130} 
  A.~Sen,
  ``Riding Gravity Away from Doomsday,''
  arXiv:1503.08130 [hep-th].
  %%CITATION = ARXIV:1503.08130;%%
  
\bibitem{9902189} 
  L.~M.~Krauss and G.~D.~Starkman,
  ``Life, the universe, and nothing: Life and death in an ever expanding universe,''
  Astrophys.\ J.\  {\bf 531}, 22 (2000)
  [astro-ph/9902189].
  %%CITATION = ASTRO-PH/9902189;%%  

\bibitem{1502.01589} 
  \changek{P.~A.~R.~Ade {\it et al.}  [Planck Collaboration],
  ``Planck 2015 results. XIII. Cosmological parameters,''
  arXiv:1502.01589 [astro-ph.CO].}
  %%CITATION = ARXIV:1502.01589;%%

\bibitem{0711.3417} 
  W.~Fischler, S.~Paban, M.~Zanic and C.~Krishnan,
  ``Vacuum bubble in an inhomogeneous cosmology: A Toy model,''
  JHEP {\bf 0805}, 041 (2008)
  [arXiv:0711.3417 [hep-th]].
  %%CITATION = ARXIV:0711.3417;%%
 
\bibitem{0908.2757} 
  D.~Simon, J.~Adamek, A.~Rakic and J.~C.~Niemeyer,
  ``Tunneling and propagation of vacuum bubbles on dynamical backgrounds,''
  JCAP {\bf 0911}, 008 (2009)
  [arXiv:0908.2757 [gr-qc]].
  %%CITATION = ARXIV:0908.2757;%%

\bibitem{0509268} 
  J.~S.~Heyl,
  ``The long-term future of space travel,''
  Phys.\ Rev.\ D {\bf 72}, 107302 (2005)
  [astro-ph/0509268].
  %%CITATION = ASTRO-PH/0509268;%%

\bibitem{1102.0007} 
  A.~Loeb,
  ``Cosmology with Hypervelocity Stars,''
  JCAP {\bf 1104}, 023 (2011)
  [arXiv:1102.0007 [astro-ph.CO]].
  %%CITATION = ARXIV:1102.0007;%%

\bibitem{1411.5022} 
  J.~Guillochon and A.~Loeb,
  ``The Fastest Unbound Stars in the Universe,''
  arXiv:1411.5022 [astro-ph.GA].
  %%CITATION = ARXIV:1411.5022;%%

\bibitem{1411.5030} 
  A.~Loeb and J.~Guillochon,
  ``Observational Cosmology With Semi-Relativistic Stars,''
  arXiv:1411.5030 [astro-ph.CO].
  %%CITATION = ARXIV:1411.5030;%%

\bibitem{0204249} 
  K.~Nagamine and A.~Loeb,
  ``Future evolution of nearby large scale structure in a universe dominated by a cosmological constant,''
  New Astron.\  {\bf 8}, 439 (2003)
  [astro-ph/0204249].
  %%CITATION = ASTRO-PH/0204249;%%

\bibitem{0706.3201} 
  T.~Clifton, S.~Shenker and N.~Sivanandam,
  ``Volume Weighted Measures of Eternal Inflation in the 
  Bousso-Polchinski Landscape,''
  JHEP {\bf 0709}, 034 (2007)
  [arXiv:0706.3201 [hep-th]].
  %%CITATION = ARXIV:0706.3201;%%
  
\bibitem{0712.1397} 
  M.~Dine, G.~Festuccia, A.~Morisse and K.~van den Broek,
  ``Metastable Domains of the Landscape,''
  JHEP {\bf 0806}, 014 (2008)
  [arXiv:0712.1397 [hep-th]].
  %%CITATION = ARXIV:0712.1397;%%
  
\bibitem{0907.4153} 
 \changek{ D.~N.~Page,
  ``Possible Anthropic Support for a Decaying Universe: 
  A Cosmic Doomsday Argument,''
  arXiv:0907.4153 [hep-th].}
  %%CITATION = ARXIV:0907.4153;%%

\bibitem{1010.5241} 
\changek{  A.~R.~Brown and A.~Dahlen,
  ``Giant Leaps and Minimal Branes in Multi-Dimensional Flux Landscapes,''
  Phys.\ Rev.\ D {\bf 84}, 023513 (2011)
      [arXiv:1010.5241 [hep-th]].}
  %%CITATION = ARXIV:1010.5241;%%

\end{thebibliography}
\end{document}